\documentclass[twocolumn,twoside]{aastex701}
\usepackage{appendix}
\usepackage{tabularx}
\usepackage{amsmath, amssymb}
\usepackage{appendix}
\usepackage{hyperref}
\usepackage{booktabs}
\usepackage{longtable}

\usepackage{multirow}
\usepackage{graphicx}
\usepackage{float} 
\usepackage{afterpage}
\usepackage{caption}

\usepackage{capt-of}

\newcommand{\tes}{\textit{TESS}}
\newcommand{\og}{\textit{OGLE}}
\newcommand{\kep}{\textit{Kepler}}

\newcommand{\jktabs}{\textsc{jktabsdim}}

\newcommand{\ravespan}{\textsc{ravespan}}

\newcommand{\isp}{\textsc{ispec}}

\newcommand{\kms}{\,km\,s$^{-1}$}
\newcommand{\U}{\textit{U}}
\newcommand{\B}{\textit{B}}
\newcommand{\V}{\textit{V}}
\newcommand{\R}{\textit{R}}
\newcommand{\I}{\textit{I}}
\newcommand{\J}{\textit{J}}

\newcommand{\K}{\textit{K}}
\newcommand{\Msun}{M$_{\odot}$}

\newcommand{\spectrumgrid}{\textsc{SPECTRUM\ ATLAS9.Castelli}}
\newcommand{\bstargrid}{\textsc{TLUSTY\ BSTAR2006}}
\newcommand{\ostargrid}{\textsc{TLUSTY\ OSTAR2002}}

\newcommand{\gaia}{\textit{Gaia}}

\newcommand{\beq}{\begin{equation}}
\newcommand{\eeq}{\end{equation}}

\begin{document}

\title{Precise physical and kinematical parameters of massive eclipsing binaries in the Small Magellanic Cloud}

\correspondingauthor{Ö. Çakırlı}
\email{omur.cakirli@ege.edu.tr}

\author[sname=Kaya, gname=M.]{M. Kaya}
\affiliation{Ege University, Science Faculty, Astronomy and Space Science Department, 35100 Bornova, İzmir, Türkiye}
\email{mehmetbilsem@gmail.com}  

\author[orcid=0000-0002-8814-8303, sname=Çakırlı, gname=Ö.]{Ö. Çakırlı}\thanks{E-mail: \href{mailto:omur.cakirli@ege.edu.tr}{omur.cakirli@ege.edu.tr} (Corresponding author)}
\affiliation{Ege University, Science Faculty, Astronomy and Space Science Department, 35100 Bornova, İzmir, Türkiye}
\email{omur.cakirli@ege.edu.tr}

\author[orcid=0000-0001-8332-522X,sname=Hoyman, gname=B.]{B. Hoyman}
\affiliation{Ege University, Science Faculty, Astronomy and Space Science Department, 35100 Bornova, İzmir, Türkiye}
\email{}

\author[orcid=0000-0003-4820-3950,sname=Özdarcan, gname=O.]{O. Özdarcan}
\affiliation{Ege University, Science Faculty, Astronomy and Space Science Department, 35100 Bornova, İzmir, Türkiye}
\email{}

\author[orcid=0000-0002-5657-6194,sname=Yontan, gname=T.]{T. Yontan}
\affiliation{Istanbul University, Faculty of Science, Department of Astronomy and Space Sciences, 34119, Beyazıt, Istanbul, Türkiye}
\email{talar.yontan@istanbul.edu.tr}

\author[orcid=0000-0003-2575-9892,sname=Canbay, gname=R.]{R. Canbay}
\affiliation{Akdeniz University, Faculty of Science, Department of Space Sciences and Technologies, 07058, Antalya, Türkiye}
\email{rmzycnby@gmail.com}

\begin{abstract}

The formation of massive stars serves an important field of study, requiring research 
into whether these stars need to form together within clusters of low-mass stars or if they 
are able to form independently. Currently, there has been an increase in the number of massive 
stars found far from clusters in the Milky Way (MW) and Small Magellanic Cloud (SMC). The 
existence of isolation and composition in binary systems is an uncommon phenomenon among 
massive stars, and their rare occurrence presents valuable insights that are regarded as 
evidence in this field. This research looks for ways to increase the catalogue of identified 
massive stars in eclipsing binaries, particularly those that show significant indications of 
isolation. For the definition of the first step, we are analysing the eight double-lined 
spectroscopic binaries in the SMC, as detailed in the OGLE variable star catalogues, with 
the goal of identifying their astrophysical properties. Later, we examine the expected 
locations and characteristics of binary stars thought to be part of a cluster, using 
absolute parameters and a cross-validation approach to determine the connection between 
kinematic analysis and these parameters. If the kinematic connection with the cluster and 
the physical parameters match precisely with the cluster's isochrone, such stars could 
act as reliable standard candles for determining the cluster's distance and age.
\end{abstract}



\section{Introduction}
\label{sec:intro}
The prevailing view is that stellar clusters are typically not formed in isolation; 
rather, they represent the densest collections of larger stellar aggregates and complexes \citep{2011EAS....51...45E} 
within a hierarchical framework of stellar structures that extends to galactic scales 
\citep{2010A&A...515A..56G,2015MNRAS.448.1847H}. Massive field stars, in contrast, are not found in clusters but 
rather appear to exist in apparent isolation, whereas cluster stars are formed together in large aggregates 
simultaneously; field stars have multiple origins. 

\citet{1957PASP...69...59R} was the first to investigate whether all massive stars are formed in clusters or if 
a considerable number may instead form in isolation as field stars. Furthermore, \citet{2003ARA&A..41...57L} and 
\citet{2010ARA&A..48..431P} indicated that stellar formation in clusters demonstrates variability in size, mass, 
and stellar composition. Such variation ranges from small, compact groups of protostars that remain embedded in 
their primordial stellar-formation regions to large young massive clusters containing approximately 10$^4$\,\Msun~of mass. Some 
of these massive field stars may only begin to scratch the surface of understanding small groups of physically 
associated stars, while others appear to be \texttt{runaway} stars that have been dynamically ejected from 
clusters. Studies that are considered to be the most significant on this topic concentrate on the origins 
of field stars.  The research conducted by \citet{2012ASPC..465..431O} studied the origins of field stars, which 
account for around 20–25\% of the massive stars that are found in galaxies and distinguish between in situ 
formations and runaway forms. While stars that are removed from their clusters due to dynamic interactions or 
supernova explosions in binary systems are known as runaway stars, \citet{2012ASPC..465..431O} showed strong 
evidence that some massive stars can indeed form in situ within small groups or even in isolation. This is 
in contrast to the traditional viewpoint, which implies that massive stars must form in massive clusters. The 
study by \citet{2016ApJ...817..113L} was particularly important in this regard. A comprehensive spectroscopic 
investigation of OB-type stars in the Small Magellanic Cloud (SMC) was carried out, focusing on the influence 
of the stellar formation environment on binarity. In the end, they considered the potential that the ratio 
of binary stars in isolated formations might be less or exhibit different period distributions in comparison 
to stars found within a cluster. 

Furthermore, \citet{2012Sci...337..444S} provided statistical evidence that the supernova 
scenario, initially suggested by \citet{1961BAN....15..265B}, is not just an infrequent occurrence but rather 
an essential part of the evolution of massive stars. This, along with observations conducted in environments 
like the SMC or Tarantula Nebula, clearly elucidates how escaped stars rise to the galactic halo and the reasons 
behind some being classified as isolated fast rotator stars
\citep{2012A&A...546A..73H,2013msao.confE..26S,2013A&A...550A.107S}.

While there may be high-mass stars that formed in true isolation, if they exist, such occurrences are likely to be extremely rare. Our study aims to identify massive eclipsing binaries in the SMC and investigate their origin via photometric and spectroscopic observations. We also get the benefit of kinematic analysis of the identified binaries to provide a wider perspective on the problem. Such a study could reveal whether they formed in an isolated environment or in a stellar cluster or association. Our research is structured as 
follows: \S\ref{sec:targ} presents the observational data sources used for modelling our systems 
\S\ref{sec:anal} presents a comprehensive overview of the methodology used to derive radial velocity 
from spectroscopic data. The approach taken to resolve both the radial velocity and light curves of 
the systems is outlined. Furthermore, it outlines the procedure to determine the atmospheric parameters, 
analysing the membership of target stars in the NGC\,290 cluster, and analysing their detailed 
kinematics. \S\ref{sec:evo}  outlines the carried out solutions for each system and evaluates 
their evolutionary stages. \S\ref{sec:sum_disc} sets out our conclusions.

\section{Observations}
\label{sec:targ}
%
%
%
%
%
\begin{table*}
\caption{\label{tab:targ} Basic information for the OGLE-SMC eclipsing binaries in the NGC\,290 cluster considered 
in this work. The stars are listed in order of increasing orbital period.}
\renewcommand{\arraystretch}{1.1}
\begin{tabular}{@{}lccccc@{}}
\hline
\hline
Target (\og)  & RA      & DEC    & $V$  & Orbital   & Spectral \\
& (J2000) & (J2000)&[mag] & period (d)& type \\
\hline
SMC-ECL-2194 & 00:51:35.8 & -73:12:45.3 & 17.06 & 1.3029 & B$^{a}$     \\
SMC-ECL-2198 & 00:51:37.2 & -73:05:50.7 & 17.44 & 1.3298 & B$^{a}$     \\
SMC-ECL-2096 & 00:51:16.8 & -73:13:02.3 & 15.94 & 1.8089 & B$^{a}$     \\
SMC-ECL-2073 & 00:51:11.4 & -73:05:21.7 & 16.78 & 1.9394 & B1+B1$^{b}$ \\
SMC-ECL-2213 & 00:51:40.3 & -73:13:20.8 & 15.36 & 2.3238 & B0$^{c}$    \\
SMC-ECL-1947 & 00:50:40.5 & -73:12:56.2 & 16.12 & 2.4013 & B3\,V$^{a}$ \\
SMC-ECL-1903 & 00:50:30.2 & -73:07:38.7 & 16.91 & 3.9426 & B3\,V$^{a}$ \\
SMC-ECL-1105 & 00:46:50.4 & -73:07:30.3 & 15.84 & 9.0576 & --          \\
\hline
\end{tabular}
\begin{flushleft}
\textit{Note: All targets are massive, early-type stars, as inferred from photometry.} References: 
$^{(a)}$\citet{2006SASS...25...47W}; $^{(b)}$\citet{2005MNRAS.357..304H}; $^{(c)}$\citet{1988ApJ...331..294H}. 
\end{flushleft}
\end{table*}

\subsection{Photometry}
\label{sec:ogle}
In this study, we used \og\ photometry from the \og-II \citep{1997AcA....47..319U}, \og-III \citep{2003AcA....53..291U}, 
and \og-IV \citep{2015AcA....65....1U} databases. The data was collected utilising the 1.3\,m Warsaw telescope at Las 
Campanas Observatory in Chile, operated by the Carnegie Institution for Science. The photometric data collected from \og\ 
utilises filters comparable to the standard $V$ and $I$ photometric bands; for instance, the \og\,$I$ band, centred at 
8\,000\,\AA, corresponds precisely with the standard Kron-Cousins $I$ band filter. The \og\,$V$ band, centred at 
5\,500\,\AA, shows minor variations compared to the standard Johnson\,$V$ band \citep{2015AcA....65....1U}.

We investigated eight binary systems, as previously defined by \citet{2016AcA....66..421P} and 
listed in Table\,\ref{tab:targ}, which includes the \og\,IDs of the objects. The \og\ online 
database\footnote{\url{https://ogledb.astrouw.edu.pl/ogle/}} provides public access to photometric data. Among 
the target stars presented in Table\,\ref{tab:targ}, only TIC\,181042539 (SMC-ECL-2213) had \tes\ \citep[Transiting 
Exoplanet Survey Satellite,][]{2015JATIS...1a4003R} data available and used for the analysis.

\subsection{Spectroscopy}
\label{spectro}
This research uses archival ESO spectroscopic database (Science Archive 
Facility\footnote{\url{http://archive.eso.org/wdb/wdb/adp/phase3_main/form}}) gathered with the multi-object 
spectrograph FLAMES, which includes the GIRAFFE spectrograph. The data was obtained via programs 072.D-0577(A), 
092.D-0110(A), 072.A-0474(A), 079.D-0562(A), 190.D-0237(A), 094.D-0056(A), 097.D-0150(A), 098.D-0876(A), and 
106.21CY.001. FLAMES functions as the multi-object, intermediate, and high-resolution spectrograph of the Very 
Large Telescope (VLT), installed at the UT2 (Kueyen telescope) in Paranal, Chile. The spectra were 
acquired with the GIRAFFE \texttt{LR02} setup along with its corresponding \texttt{LR02} setting, which provides 
a spectral resolving power of $\lambda/\delta\lambda$ $\sim$ 6500. This resolving power is adequate to identify 
stars where helium lines might be influenced by nebular contamination, as established through inspection of 
\texttt{Balmer} lines. The \texttt{LR02} data offer continuous wavelength coverage from 3964 to 4567\,\AA, allowing 
access to multiple spectral lines that are ideal for precise radial velocity measurements. In this GIRAFFE 
configuration, observations of 8 stars resulted in the collection of a total of 127 spectra. The details of 
these observations are presented in Table\,\ref{tab:spectra_summary}.

The essential step in analysing the obtained spectra is the normalisation process. This is evident in studies 
on stellar abundances \citep[e.g.,][]{2014A&A...569A.111B,2015A&A...577A..67S,2016A&A...592A..87A} and in the 
analysis of low-mass stellar atmospheres. Later, entails cosmic correction, which is equally important as 
normalisation.

\begin{table}
\caption{\label{tab:spectra_summary}\textbf{ Summary of spectroscopic observations.}}
\fontsize{7.0}{8.0}\selectfont
\setlength{\tabcolsep}{3.0pt}
\renewcommand{\arraystretch}{1.15}
\centering
\begin{tabular}{@{}lcccc@{}}
\hline
\hline
Target (\og) & Observation Dates & Exposure (s) & SNR & $N_{\rm Giraffe}$ \\
\hline
SMC-ECL-2194 & 2003/11/16-23 & 2\,595 &  35.1 & 16 \\
SMC-ECL-2198 & 2003/11/16-25 & 2\,595 &  29.4 & 16 \\
SMC-ECL-2096 & 2003/11/16-23 & 2\,600 &  57.9 & 16 \\
SMC-ECL-2073 & 2003/11/16-23 & 2\,595 &  66.2 & 16 \\
SMC-ECL-2213 & 2003/11/16-23 & 2\,700 &  71.2 & 16 \\
SMC-ECL-1947 & 2003/11/16-25 & 2\,700 &  54.7 & 21 \\
SMC-ECL-1903 & 2003/11/16-23 & 2\,595 &  40.1 & 21 \\
SMC-ECL-1105 & 2003/10/02-25 & 2\,700 &  61.8 & 5  \\
\hline
\end{tabular}
\end{table}
%
%
%

%
%
\begin{figure*}
\includegraphics[width=0.99\textwidth]{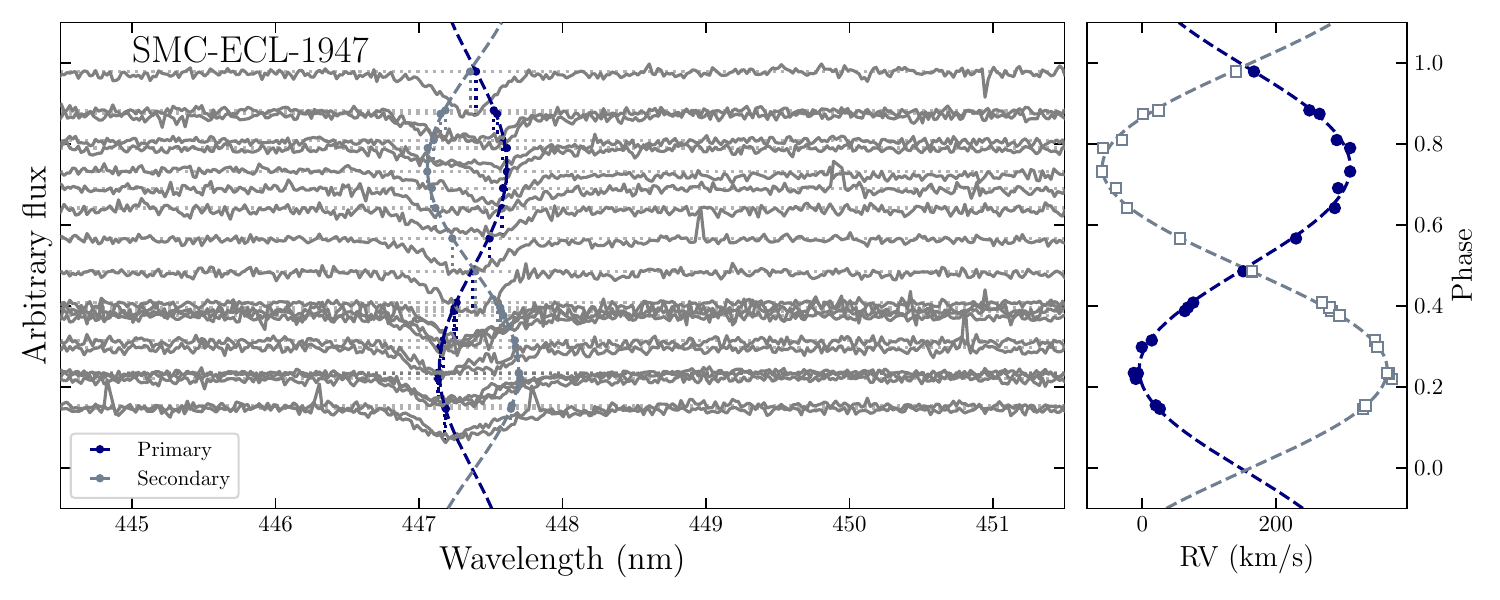}
\caption{The left panel shows a set of absorption lines involving a secondary component, which 
matches the orbital period cycle of the system. The variation of the He\,{\sc i}\,$\lambda$4471 
\AA\ line profiles is represented by dashed lines. The right panel displays the radial velocity of 
the absorption lines. The filled circles on the radial velocity plot indicate the primary, whereas 
the open squares represent the secondary.}
\label{phased_mcmc_rv}
\end{figure*}
%
%

The RVs for each object were established by using cross-correlation of the observed spectra with a 
synthetic template, constructed by combining synthetic spectra that correspond to both the 
primary and secondary components. We adopt the Cross-Correlation 
Function \citep[CCF;][]{1979AJ.....84.1511T} method to detect the RV signals of the 
components. This method allowed us to detect signals from these 
components. In practise, we utilise the \ravespan\ code \citep[Radial Velocity and Spectrum Analyser;]
[]{2017ApJ...842..110P} to apply the CCF method. The template spectra used in the CCF method were specifically created for practical measurements of RVs in stars with effective temperatures between 10\,000 and 30\,000\,K, in increments of 3\,000 K. Table\,\ref{table_radial_velocities} presents the velocity values 
for both components of our SB2 systems, including their associated errors and the heliocentric 
Julian dates of the observations.

\section{Analysis methods}
\label{sec:anal}
\subsection{Atmospheric parameters}
\label{sec:atm}
We used a \textsc{python} code that makes use of the \isp\ routines of 
\citet{2014A&A...569A.111B,2019MNRAS.486.2075B} to generate model spectra in order 
to determine the atmospheric parameters of each component star in all 8 of our 
binary systems. Depending on the effective temperature, one of the 
\spectrumgrid\ grid, \bstargrid\ grid, or \ostargrid\ grid was used to generate the 
synthetic spectra.

The code optimises up to ten stellar parameters for each component simultaneously. These 
parameters include surface temperature ($\rm T_{eff\,1,2}$), surface gravity ($\log g_{1,2}$), 
metallicity ($[M/H]_{1,2}$), alpha enhancement ($[\alpha Fe]_{1,2}$), micro-turbulent velocity 
($\nu_{mic,1,2}$), macro-turbulent velocity ($\nu_{mac,1,2}$), projected rotational velocity 
($v \sin i_{1,2}$), limb darkening coefficients ($LD_{1,2}$), radial velocity ($RV_{1,2}$), and 
the fractional light contribution of each component to the total light ($l_{frac_{1,2}}$).

The individual spectra of each component were synthesized with the respective parameter 
set and scaled according to the light contribution, shifted by the radial velocity value. These 
spectra were then combined to form a composite spectrum which was then compared with the observed 
spectrum. This is an iterative process to hone in on the parameter set to get it right. This 
synthetic composite spectrum can cover the whole observed spectrum or be limited to certain 
\emph{segments} to reduce the computational time and to improve the quality of the solutions. For 
a correct determination of the expected atmospheric parameters, attention should be paid to certain 
lines in the spectral range of the stars selected for the analysis. We used the wavelength range 
400 to 450 nm, in which the spectral lines of both components are visible.

The solution involves fixing certain parameters, such as $\nu_{mic}$, $\nu_{mac}$, and $LD$, while 
establishing relationships for others, including $[M/H]_1 = [M/H]_2$ and $l_{frac_2} = 1 - l_{frac_1}$. This 
method limits the solution according to these defined links. Furthermore, it is possible to define 
variance ranges for independent parameters, which enables the solution to be restricted within 
these specified ranges. The initial parameter-seeking setup utilises broad ranges and substantial 
step sizes to guarantee the identification of the global minimum while minimising computational 
time. Nonetheless, differentiating between micro- and macro-turbulent velocities, as well as 
rotational velocity, poses a significant challenge. This challenge necessitates spectra with 
exceptionally high resolution and a strong signal-to-noise ratio (SNR) to effectively tackle 
the degeneracy among these parameters \citep{2014MNRAS.444.3592D}. To address this degeneracy 
and create a standard procedure for analysing spectra with lower resolution and SNR, we adopt 
$\nu_{mic}$ as 2\,\kms~ to align with the grid we are using. Our initial analyses have shown 
extremely high projected rotational velocities, which could greatly diminish the effect of 
macro-turbulence in spectral lines; therefore, we have set $\nu_{mac}$ to zero.

\begin{figure}
\includegraphics[width=0.5\textwidth]{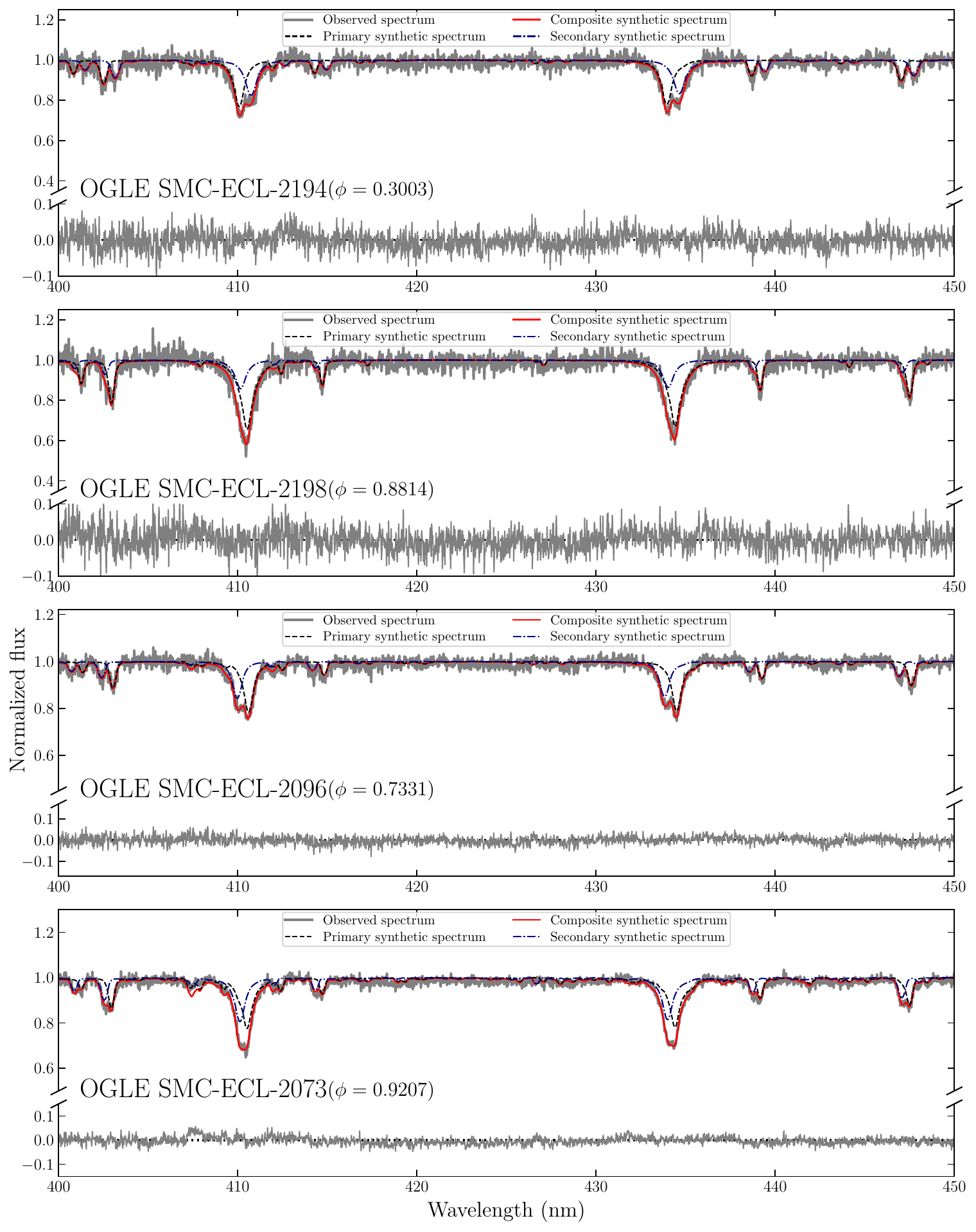}
\includegraphics[width=0.5\textwidth]{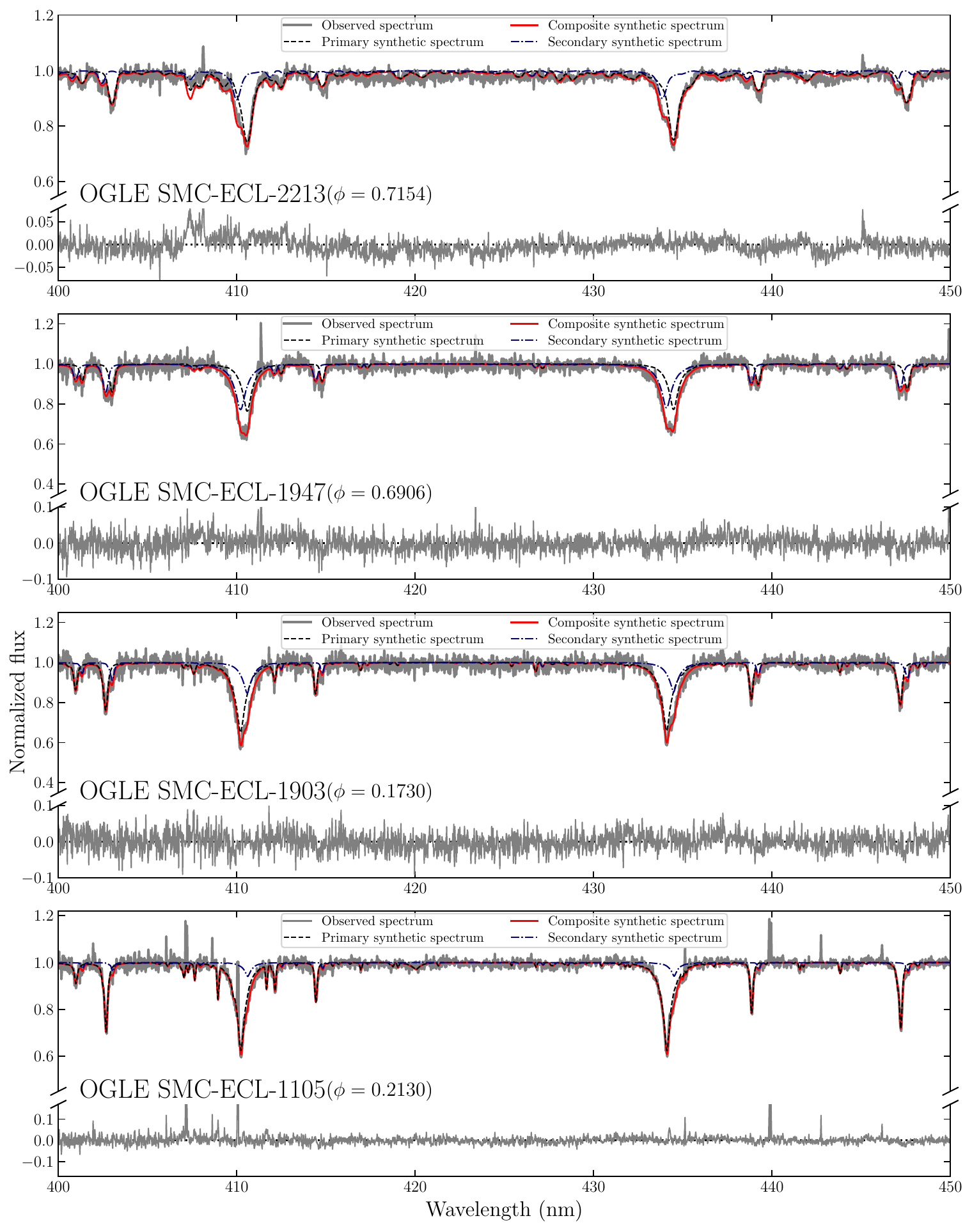}
\caption{The composite spectra of the systems are compared to a combined synthetic model, which is 
represented by the dashed line for the primary and the dot-dashed line for the secondary star. This model 
is created from the models that match each system the best.}
\label{atm}
\end{figure}

The \textsc{emcee} package \citep{emcee_2013PASP}, a realistic implementation of the MCMC method, was 
used to identify the best-fitting parameters and their uncertainties. First, we set the [M/H] parameter 
as well as the effective temperature and $\log g_{1,2}$ parameters for each component to be configurable. We 
also included $v\sin i$ parameters as modifiable if we saw rotational broadening in earlier trials. The 
initial fixed parameters for light curve modelling were the T$_{1}$ and [M/H] parameters that were determined 
at the end of the MCMC run. After that, we computed the target system's absolute attributes by modelling 
light curves. For the next MCMC run, we fixed the $\log g_{1,2}$ values of the components among these 
attributes. By successfully constraining the $\log g_{1,2}$ values and resolving any degeneracy among the 
remaining changeable parameters, this approach produces atmospheric parameter estimates that are more 
accurate.
Graphical representations of the results of the MCMC analysis and 
the consistency between the observed and fitted model spectra for the systems are presented in 
Figure\,\ref{atm}. Table\,\ref{tab:atmosphericParam} contains a tabulation of the numerical results. The 
reported uncertainties, particularly for the effective temperatures, seem too low (less than 130\,K) for 
several systems given the resolution and SNR of the measured spectra as well as the spectroscopic 
characteristics of the target systems. The real uncertainties, which are mostly caused by ambiguity 
in the continuum level, are expected to be approximately 150\,K. 

%

%
%
%

%
%
%
%

%
%
\begin{table}
\fontsize{6.5}{9.2}\selectfont
\setlength{\tabcolsep}{2.7pt}
\caption{Atmospheric analysis results of the target systems. We adopted the same metallicity for both components. }
\label{tab:atmosphericParam}
\centering
\begin{tabular}{lccccc}
\hline
\hline
Catalog~IDs                   &T$\rm_{eff}$ &$\log g^a$  &[M/H]                       &$v\sin i_{\rm obs}$ & \multirow{2}{*}{Reduced $\chi^2$} \\
                              &(K)          &(dex)     &(dex)                       &(km~s$^{-1}$)       &                                     \\
\hline                                                    
\multirow{2}{*}{SMC-ECL-2194} &29\,050$^{+30}_{-50}$ &4.31 &\multirow{2}{*}{-1.198$^{+0.022}_{-0.033}$}  &213$^{+7}_{-11}$&\multirow{2}{*}{1.6021}\\
                              &26\,450$^{+20}_{-70}$ &4.32 &                            &177$^{+6}_{-15}$                 &                       \\ 
\multirow{2}{*}{SMC-ECL-2198} &26\,300$^{+17}_{-33}$ &3.97 &\multirow{2}{*}{-1.009$^{+0.013}_{-0.031}$}  &172$^{+3}_{-9}$ &\multirow{2}{*}{2.0129}\\
                              &19\,290$^{+40}_{-52}$ &4.10 &                            &125$^{+4}_{-9}$                  &                       \\ 
\multirow{2}{*}{SMC-ECL-2096} &31\,250$^{+23}_{-83}$ &3.80 &\multirow{2}{*}{-1.004$^{+0.009}_{-0.012}$}  &174$^{+8}_{-17}$&\multirow{2}{*}{2.0011}\\
                              &20\,400$^{+19}_{-93}$ &3.66 &                            &191$^{+6}_{-13}$&                                        \\ 
\multirow{2}{*}{SMC-ECL-2073} &30\,000$^{+32}_{-103}$&3.77 &\multirow{2}{*}{-0.267$^{+0.009}_{-0.013}$}  &144$^{+5}_{-9}$ &\multirow{2}{*}{2.1102}\\
                              &20\,100$^{+41}_{-121}$&3.48 &                            &165(13)         &                                        \\ 
\multirow{2}{*}{SMC-ECL-2213} &28\,510$^{+26}_{-91}$ &3.70 &\multirow{2}{*}{0.089$^{+0.041}_{-0.044}$}   &212$^{+4}_{-8}$ &\multirow{2}{*}{1.1255}\\
                              &18\,210$^{+44}_{-101}$&3.71 &                            &179$^{+7}_{-11}$&                                        \\ 
\multirow{2}{*}{SMC-ECL-1947} &12\,280$^{+22}_{-38}$ &3.81 &\multirow{2}{*}{-0.489$^{+0.032}_{-0.033}$}  &117$^{+7}_{-17}$&\multirow{2}{*}{2.0033}\\
                              & 9\,520$^{+39}_{-77}$ &4.32 &                            &79$^{+5}_{-11}$ &                                        \\ 
\multirow{2}{*}{SMC-ECL-1903} &19\,400$^{+24}_{-67}$ &3.98 &\multirow{2}{*}{0.193$^{+0.022}_{-0.031}$}   &81$^{+3}_{-9}$  &\multirow{2}{*}{2.0012}\\
                              &17\,800$^{+61}_{-111}$&4.02 &                            &75$^{+6}_{-13}$ &                                        \\ 
\multirow{2}{*}{SMC-ECL-1105} &32\,300$^{+24}_{-87}$ &3.90 &\multirow{2}{*}{-0.877$^{+0.019}_{-0.025}$}  &52$^{+2}_{-9}$  &\multirow{2}{*}{1.9441}\\
                              &20\,725$^{+71}_{-122}$&3.89 &                            &58$^{+3}_{-10}$ &                                        \\ 
\hline
\end{tabular}\\
\scriptsize
Note $^a$: The parameter is obtained from the light curve solution and fixed during the atmospheric analysis.
\end{table}

\subsection{Light and radial velocity curve modelling}
\label{sec:rvlc}
The light and measured radial velocity curves were simultaneously modelled 
using the Wilson-Devinney (hereafter WD) software \citep{1971ApJ...166..605W,1979ApJ...234.1054W}. To 
utilise the WD code efficiently and rapidly, we employ the \textsc{PyWD2015} framework. This framework 
provides a graphical user interface for the 2015 version of the code and facilitates the simultaneous 
analysis of light and radial velocity curves \citep{WD2015_2014ApJ,PyWD2015_2020CoSka..50..535G}. The 
primary advantage of the code is that the WD program 
uses a differential correction algorithm to determine a set of parameters that characterise observables in 
the solution, while one of the modelling options, \texttt{Mode\,2}, is intended for detached eclipsing 
binaries. Another aspect to consider is \texttt{Mode\,5} of the WD code, which is used for analysing the 
light curves of semi-detached eclipsing binary stars, where the secondary star fills its limiting Roche lobe, 
and differs from \texttt{Mode\,2}. While we employed the albedo ($A$) and gravity brightening ($g$) coefficients 
for both components, these were set to 1.0, which are standard values for stars with radiative envelopes 
\citep{1980MNRAS.193...79R}. In our modelling, we utilised the linear limb darkening 
law, incorporating automatically updated coefficients from precomputed tables provided by 
\citet{1993AJ....106.2096V}, along with the stellar atmosphere formulation. 

\begin{figure*}
\includegraphics[width=0.33\textwidth]{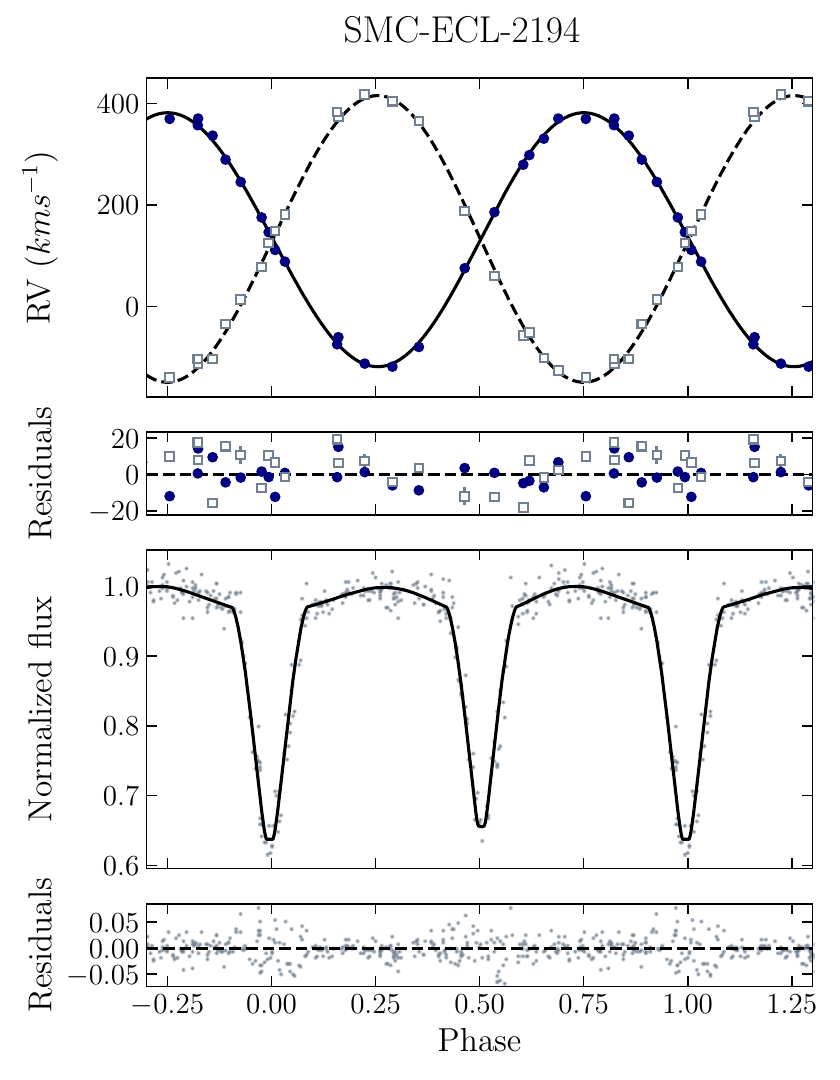}
\includegraphics[width=0.33\textwidth]{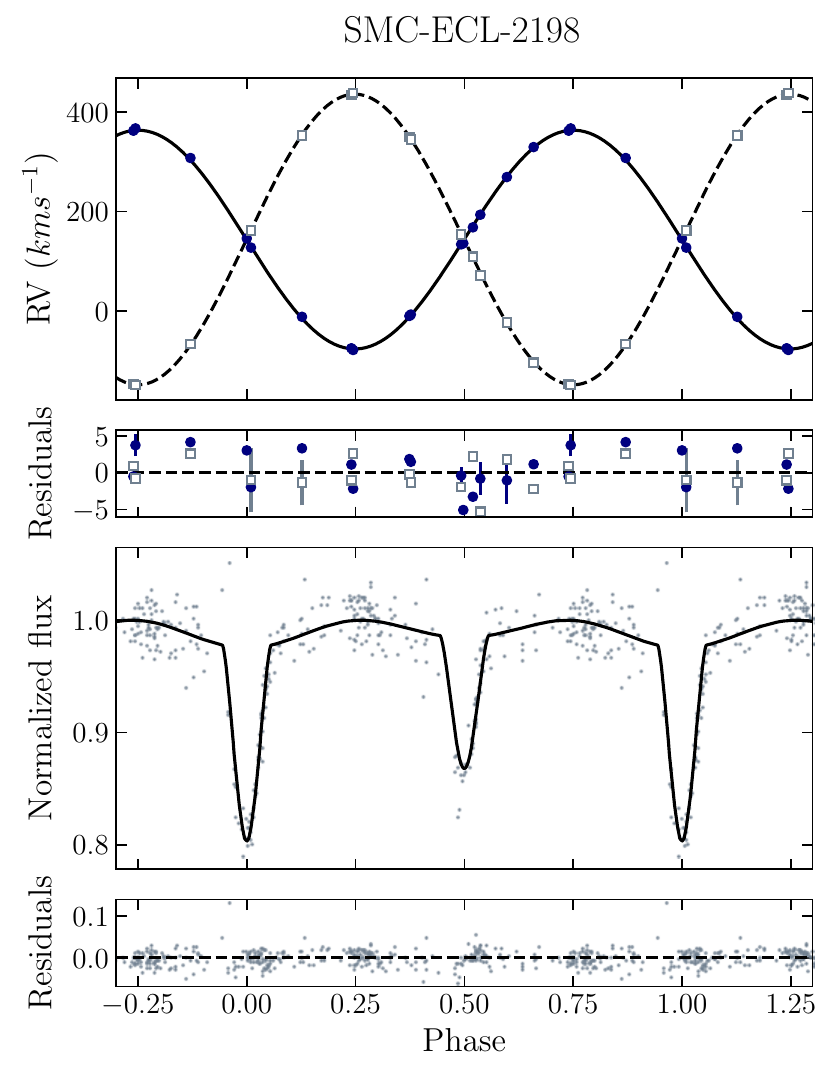}
\includegraphics[width=0.33\textwidth]{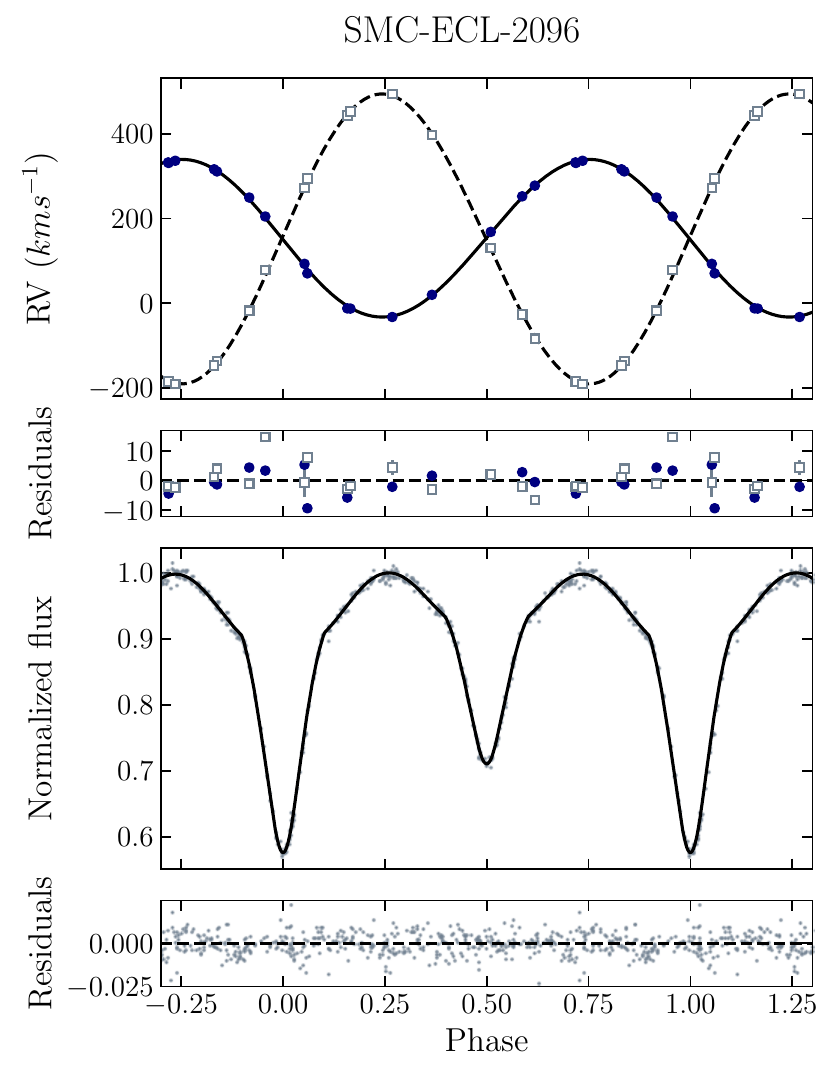}
\includegraphics[width=0.33\textwidth]{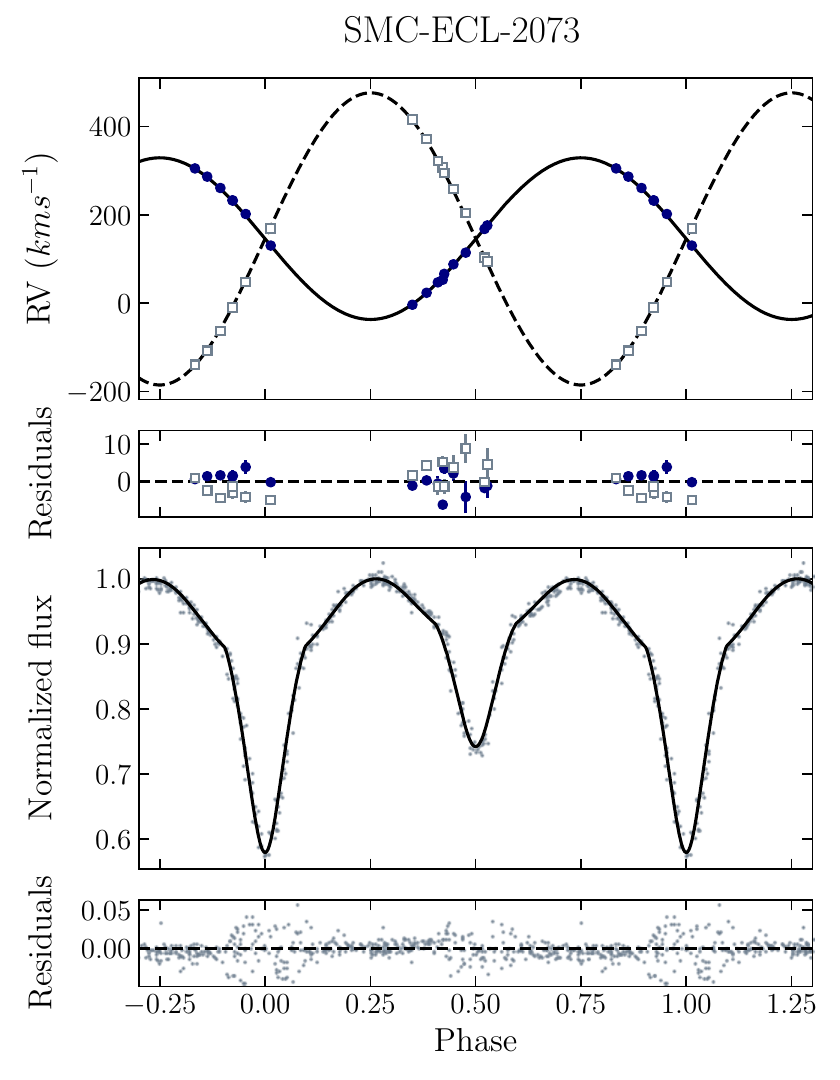}
\includegraphics[width=0.33\textwidth]{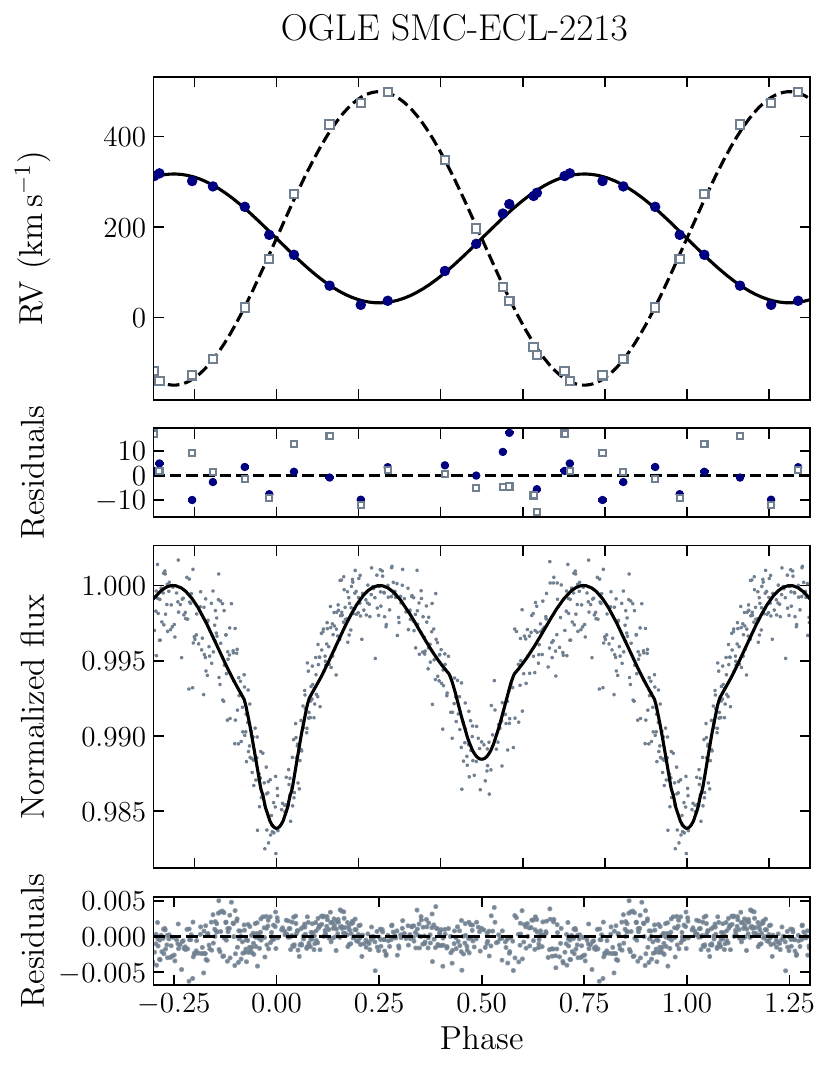}
\includegraphics[width=0.33\textwidth]{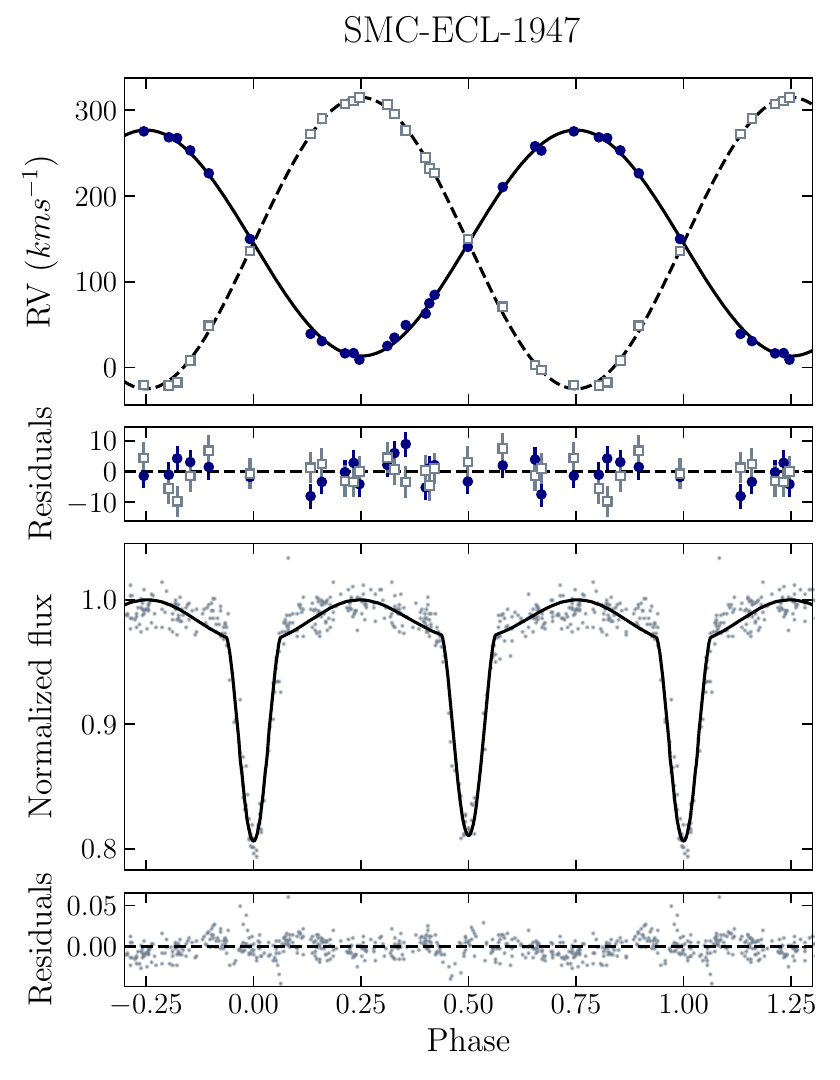}
\includegraphics[width=0.33\textwidth]{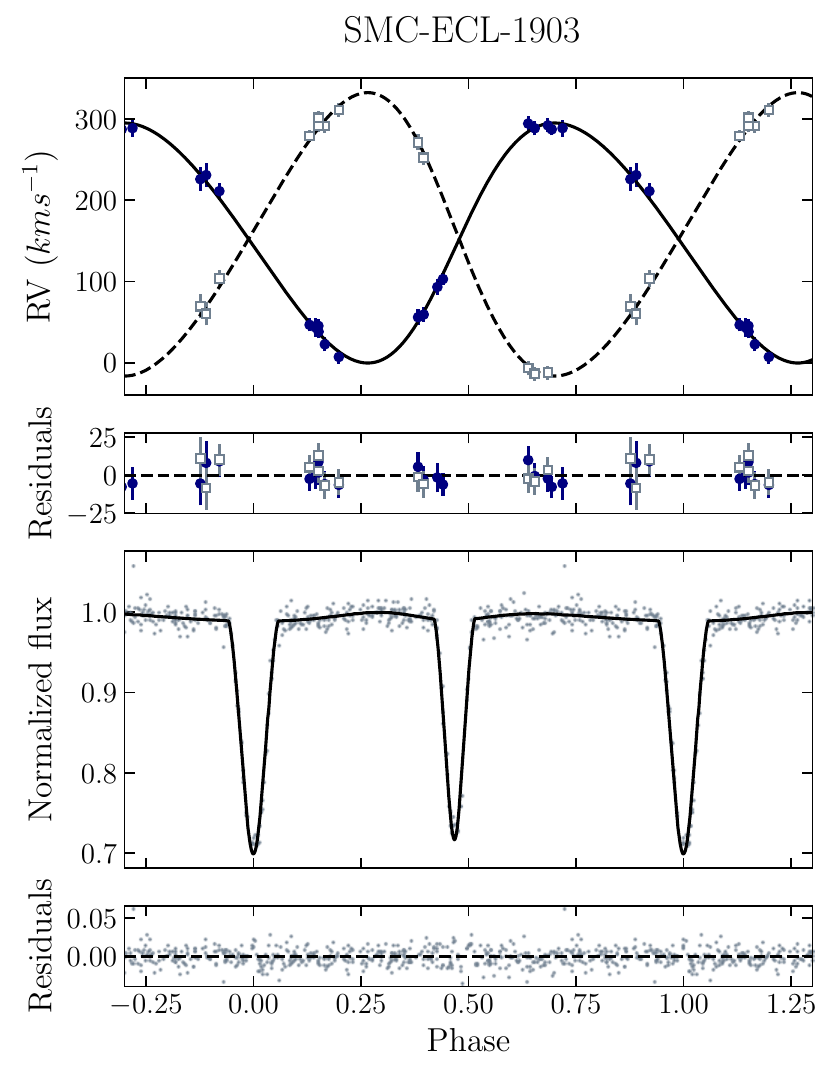}
\includegraphics[width=0.33\textwidth]{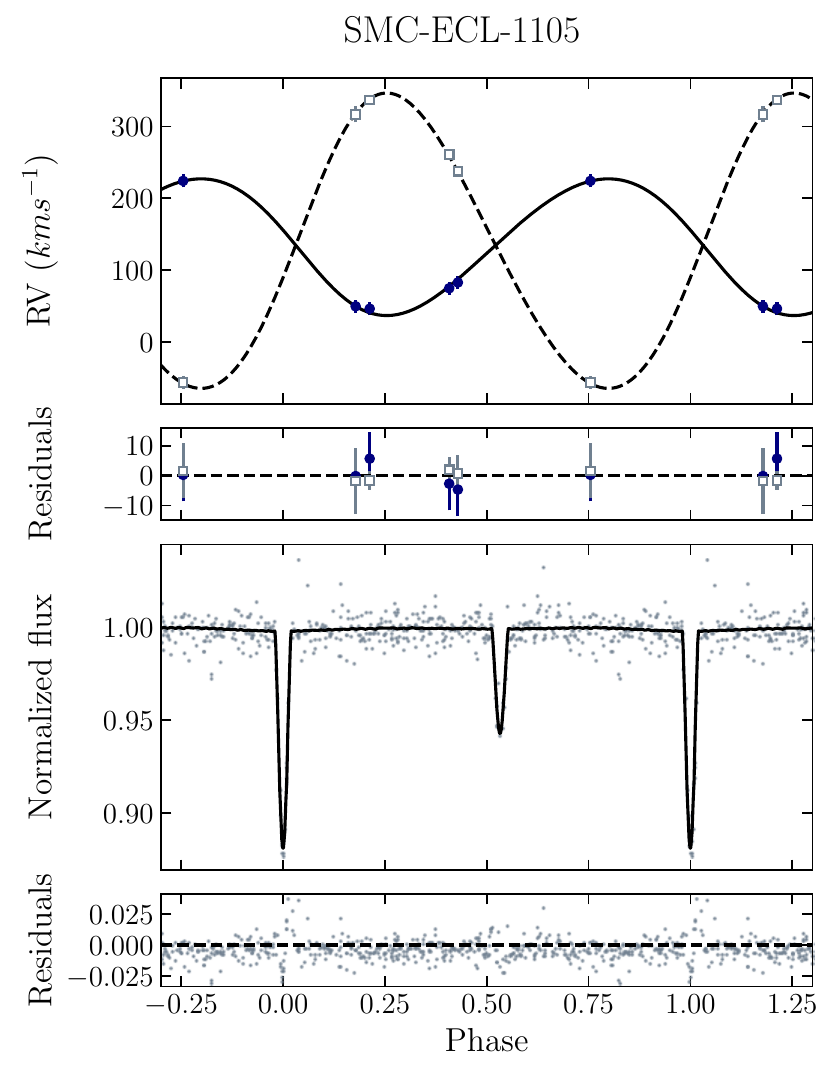}
\caption{The figure presents a graphical representation of the spectroscopic orbit and light curve models for the 
target systems. The first two panels illustrate the phase-folded spectroscopic orbit model. The continuous and dashed 
curves represent the more massive primary and the less massive secondary components, respectively. Filled circles 
represent observations for the primary component, whereas open squares indicate those for the secondary 
component, including their corresponding residuals.
}
\label{fig_LC_RV_plot1}
\end{figure*}

The initial values for the orbital parameters have been estimated based on a preliminary geometrical 
analysis of the radial-velocity curve. One of the most important initial parameters of the analysis is 
the temperature of the primary component (T$_1$), which we maintained as a constant throughout the light 
and radial velocity solution. We established the temperature of the primary star by making the assumption 
that this component is the hotter one, and we fixed the value of T$_1$ by using the data from the atmospheric 
analysis that we had previously obtained. In the same way, we made the assumption that the metallicity was 
consistent with the value that was calculated based on the analysis of the atmosphere. We initiated the 
solution process by simultaneously modifying the following parameters: eccentricity ($e$), semi-major axis 
($a$), orbital inclination ($i$), longitude of the periastron ($\omega$), dimensionless surface potentials 
($\Omega_1$, $\Omega_2$), systemic velocity ($V_{\gamma}$), mass ratio ($q$=M$_2$/M$_1$), effective 
temperature of the secondary component (T$_2$), orbital period ($P$), phase shift ($\varphi$), luminosity 
of the primary component in the specified filter (L$_1$), and third light ($\ell_3$). After reaching a 
satisfactory solution, we present the best-fitting light curves and RV curves
derived from the binary model, along with their corresponding O-C residuals, in 
Figure\,\ref{fig_LC_RV_plot1}. The results of the best-fitting models are summarised in 
Table\,\ref{salt_table1} and \ref{salt_table2}, which indicates that the convergence solutions, along 
with the associated parameter errors provided by the WD code, are underestimated. In order to overcome 
this limitation, we will compute the parameter errors, as detailed in the subsequent section.
\subsection{Absolute dimensions and distance}
\label{sec:anal:phys}
Prior to comparing the model predictions with observational data, we took extreme care in determining 
the orbital and stellar parameters of the systems. The physical parameters are obtained from models 
of light and radial velocity curves, using a \texttt{FORTRAN}-based code called \jktabs, developed by 
\citet{2005A&A...429..645S}. An approach known as perturbation analysis was used to derive the output 
error bars. The code requires 
data derived from modelling solutions, which include the period, radial velocity semi-amplitude, inclination, 
eccentricity, component temperatures, fractional radii, reddening, and the availability of photometry with 
the visual magnitudes of the system in the Johnson-Cousins \emph{UBVRI} and the Two Micron All-Sky Survey 
(2MASS) \emph{JHKs} bands. Based on the provided information, the code computes the radius, mass, and 
luminosity for each component. For calculating luminosity, the bolometric correction data sourced 
from \citet{2002A&A...391..195G} is utilised.

Measuring the distance of the systems is crucial for our understanding of their formation and evolution. In 
this context, it has been recently proposed that stellar clustering might be the key to shaping the structural 
architecture of galaxies. Assessing interstellar extinction is essential for numerous fields of astronomical 
research. Nonetheless, reddening continues to be a primary impediment to the precise measurement of interstellar 
distances. The interstellar extinction for each of our target stars was determined using two methods, as outlined 
by \citet{2023MNRAS.523.1676C}, which provides a detailed explanation comparable to this study. Additionally, the 
distances to the systems can be readily quantified.

On two methods for estimating reddening values within the context of the Magellanic 
Clouds. The first method using the reddening maps established by \citet{2011AJ....141..158H}. 
Incorporating recent maps would minimally affect findings, with adjustments of less than $-$0.003\,mag 
for \citet{2021ApJS..252...23S} and a potential increase of $+$0.008\,mag for \citet{2020ApJ...889..179G}. 
The second method encompasses effective temperatures from atmospheric analyses discussed in 
\S\,\ref{sec:atm}. Effective temperatures were employed to derive intrinsic $(V-I)$ and $(V-K)$ colours 
through relevant calibrations \citep{1998A&A...339..858D,2000AJ....119.1424H,2005ApJ...626..465R}. These 
intrinsic colours were compared to observed colours from an initial solution to calculate $E(V-I)$ and 
$E(V-K)$ colour excesses, subsequently leading to an estimated $E(B-V)$. The average reddening was 
computed based on components with estimates from both methods. The results, including averaged reddening 
values and final distances, are detailed in Tables \ref{salt_table1} and \ref{salt_table2}, which 
also showcase physical properties, distances, and associated uncertainties.

It seems that the predicted distances of the systems are in agreement with the average 
distance of the NGC\,290 cluster. Based on the \citet{2007A&A...466..165C}, the distance to the 
NGC\,290 cluster was confirmed to be 60.3\,kpc. On the other hand, the \citet{2020ApJ...904...13G} 
found the distance to the SMC to be 62.4\,kpc, reaching an accuracy of 2\%, which is dependent on 
late-type eclipsing binary stars. All of the distances that we found for the target stars are in 
agreement with the ones that were published in the research. Kinematic analysis allows for the 
verification of the positions of the stars inside the SMC as well as their membership in the 
NGC\,290 cluster.
%

\begin{table*} 
    \fontsize{7.2}{8.6}\selectfont
\centering  
\caption{Physical properties of the systems. Quantities in brackets are uncertainties in the final digits of the preceding numbers. T$_0$ is given as BJD(TDB)-24\,00000.0.}
\label{salt_table1} 
\begin{tabular}{lcccccccccc}
    \hline
    \multirow{2}{*}{Parameter}                   & \multicolumn{2}{c}{SMC-ECL-2194}  & \multicolumn{2}{c}{SMC-ECL-2198}  & \multicolumn{2}{c}{SMC-ECL-2096}  & \multicolumn{2}{c}{SMC-ECL-2073}\\
                                                 &    $primary$     &  $secondary$   &    $primary$     &  $secondary$   &    $primary$     &  $secondary$   &    $primary$     &  $secondary$ \\ 
\hline
$Mode$                                           &\multicolumn{2}{c}{\texttt{Mode~2}}&\multicolumn{2}{c}{\texttt{Mode~2}}&\multicolumn{2}{c}{\texttt{Mode~5}}&\multicolumn{2}{c}{\texttt{Mode~5}}\\%
T$_0$                                            & \multicolumn{2}{c}{57001.0595(1)} & \multicolumn{2}{c}{57001.2875(2)} & \multicolumn{2}{c}{57000.7471(2)} & \multicolumn{2}{c}{57001.5821(1)} \\ %
P (day)                                          &  \multicolumn{2}{c}{1.302942(2)}  &  \multicolumn{2}{c}{1.329851(3)}  &  \multicolumn{2}{c}{1.808929(2)}  &  \multicolumn{2}{c}{1.939408(2)}  \\ %
$a$ $sin\,i$ (R$_{\odot}$)                       &   \multicolumn{2}{c}{12.932(107)} &   \multicolumn{2}{c}{13.463(133)} &   \multicolumn{2}{c}{18.903(117)} &   \multicolumn{2}{c}{19.696(108)} \\ %
$\gamma$ (kms$^{-1}$)                            &   \multicolumn{2}{c}{160.4(9)}    &   \multicolumn{2}{c}{143.6(2)}    &   \multicolumn{2}{c}{152.7(4)}    &   \multicolumn{2}{c}{146.2(9)}    \\ %
K$_{1,2}$ (km s$^{-1}$)                          &    241.7(1.2)    & 260.8(1.3)     &    220.1(1.4)     &  292.3(1.5)   &     186.3(1.5)   &  342.6(1.9)    &    183.1(6)      &  330.9(1.7)    \\ %
$e$                                              &   \multicolumn{2}{c}{0.022(1)}    &      \multicolumn{2}{c}{0.0}      &    \multicolumn{2}{c}{0.0}        &   \multicolumn{2}{c}{0.002(2)}    \\ %
$\omega$(${\degr}$)                              &   \multicolumn{2}{c}{257.4(1.1)}  & \multicolumn{2}{c}{90}   & \multicolumn{2}{c}{90}   &   \multicolumn{2}{c}{-37.1(1.8)}  \\ %
$q$ (M$_2$/M$_1$)                                &   \multicolumn{2}{c}{0.927(5)}    &   \multicolumn{2}{c}{0.753(6)}    &   \multicolumn{2}{c}{0.544(6)}    &   \multicolumn{2}{c}{0.553(9)}    \\ %
$i$ (${\degr}$)                                  &    \multicolumn{2}{c}{75.9(7)}    &    \multicolumn{2}{c}{76.7(2)}    &    \multicolumn{2}{c}{78.8(7)}    &    \multicolumn{2}{c}{78.4(5)}    \\ %
T$_{\rm eff\,1,2}$ (K)                           & 29\,050$^a$[fix] &  27\,070(190)  & 26\,300$^a$[fix] &  20\,340(130)  & 31\,250$^a$[fix] &  23\,050(140)  & 30\,000$^a$[fix] & 21\,050(160)   \\ %
$\Omega_{1,2}$                                   &    4.8444(53)    & 4.5402(61)     &    2.6055(23)    &   3.3323(54)   &    2.4311(39)    &  2.6822(55)    &   2.3445(21)     & 2.2777         \\ %
$A_{1,2}$                                        &   \multicolumn{2}{c}{1.0[fix]}    &   \multicolumn{2}{c}{1.0[fix]}    &   \multicolumn{2}{c}{1.0[fix]}    &   \multicolumn{2}{c}{1.0[fix]}    \\ %
$g_{1,2}$                                        &   \multicolumn{2}{c}{1.0[fix]}    &   \multicolumn{2}{c}{1.0[fix]}    &   \multicolumn{2}{c}{1.0[fix]}    &   \multicolumn{2}{c}{1.0[fix]}    \\ %
$X_{1,2}^c$                                      &      0.252       &     0.263      &      0.266       &     0.302      &      0.186       &     0.204      &      0.184       &     0.212      \\ %
$l_{1}/(l_1 + l_2)$                              &    \multicolumn{2}{c}{0.51(3)}    &    \multicolumn{2}{c}{0.91(4)}    &    \multicolumn{2}{c}{0.86(2)}    &    \multicolumn{2}{c}{0.88(5)}    \\ %
$r_{\rm 1,2}$                                    &   0.2723(2)      &  0.2574(2)     &  0.4187 (1)     &   0.3153(2)     &   0.4732 (2)     &  0.4105 (1)    &    0.4653(4)     &   0.4815(4)    \\ %
$\Sigma$ $W(O-C)^2$                              &    \multicolumn{2}{c}{0.028}      &    \multicolumn{2}{c}{0.033}      &    \multicolumn{2}{c}{0.033}      &    \multicolumn{2}{c}{0.041}      \\ 
\hline
Absolute parameters                              &                  &                &                  &                &                  &                &                  &                \\ 
\hline
M$_{1,2}$ (M$_{\odot}$)                          &    9.738(11)     &   9.025(13)    &    11.471(18)    &   8.636(10)    &   19.067 (11)    &   10.361(17)   &    18.753(23)    &   10.371(29)   \\ %
R$_{1,2}$ (R$_{\odot}$)                          &     3.631(19)    &   3.434(25)    &     5.791(3)     &    4.364(3)    &     9.125(6)     &   7.917(8)     &    9.365(4)      &   9.688(7)     \\ %
$\log(g_{1,2})$ (cgs)                            &     4.31(2)      &   4.32(3)      &     3.98(2)      &    4.11(2)     &     3.80(4)      &    3.66(5)     &     3.77(1)      &    3.48(1)     \\ %
$(v_{1,2}\sin i)_{\rm calc}$ (km s$^{-1}$)$^d$   &      140(2)      &     133(4)     &      222(1)      &    169(1)      &      251(4)      &     223(7)     &      245(8)      &     249(11)    \\ %
$\log(L_{1,2}/L_{\odot})$                        &    3.924(33)     &   3.752(34)    &     4.144(1)     &    3.797(2)    &     4.773(4)     &  4.196(6)      &     4.803(2)     &    4.204(3)    \\ %
M$_{\rm bol\,1,2}$ (mag)                         &     -5.062(87)  &  -4.630(98)     &     -5.603(114)  &  -4.830(95)    &     -7.188(111)  &  -5.744(105)   &   -7.253(91)     & -5.768(98)     \\ %
(m-M)$_V$ (mag)                                  &   \multicolumn{2}{c}{19.033(4)}   &   \multicolumn{2}{c}{19.014(11)}  &   \multicolumn{2}{c}{18.889(13)}  &   \multicolumn{2}{c}{19.214(35)}  \\ %
$E(B-V)$ (mag)                                   &   \multicolumn{2}{c}{0.188(87)}   &   \multicolumn{2}{c}{0.477(112)}  &   \multicolumn{2}{c}{0.546(15)}   &  \multicolumn{2}{c}{0.167(5)}     \\ %
$d$ (kpc)$^b$                                    &   \multicolumn{2}{c}{60.1(2.3)}   &   \multicolumn{2}{c}{62.3(2.1)}   &   \multicolumn{2}{c}{63.4(1.4)}   &   \multicolumn{2}{c}{66.3(1.8)}   \\ %
\hline
\end{tabular} 
\begin{minipage}{0.95\textwidth}
\footnotesize
\emph{Notes}: 
$^a$T$_{\rm eff\,_1}$ were found in atmospheric analysis in \S~\ref{sec:atm}.
$^b$The \jktabs\ distance is calculated from 2MASS magnitudes. 
\end{minipage}
\end{table*}

\begin{table*} 
\fontsize{7.2}{8.6}\selectfont
\centering
\caption{Physical properties of the systems. Quantities in brackets are uncertainties in the final digits 
of the preceding numbers. T$_0$ is given as BJD(TDB)-24\,00000.0.}
\label{salt_table2} 
\begin{tabular}{lcccccccccc}
\hline
\multirow{2}{*}{Parameter}                     &\multicolumn{2}{c}{OGLE\,SMC-ECL-2213} &\multicolumn{2}{c}{OGLE\,SMC-ECL-1947} &\multicolumn{2}{c}{OGLE\,SMC-ECL-1903} &\multicolumn{2}{c}{OGLE\,SMC-ECL-1105} \\
                                               &     $primary$     &    $secondary$    &     $primary$     &    $secondary$    &     $primary$     &    $secondary$    &     $primary$     &    $secondary$    \\ 
\hline
$Mode$                                         & \multicolumn{2}{c}{\texttt{Mode~5}}   & \multicolumn{2}{c}{\texttt{Mode~2}}   & \multicolumn{2}{c}{\texttt{Mode~2}}   & \multicolumn{2}{c}{\texttt{Mode~2}}   \\%
T$_0$                                          & \multicolumn{2}{c}{57001.8070(2)}     & \multicolumn{2}{c}{57002.5713(1)}     & \multicolumn{2}{c}{57001.2523(2)}     & \multicolumn{2}{c}{57009.5280(2)}     \\%
P (day)                                        & \multicolumn{2}{c}{2.323886(2)}       & \multicolumn{2}{c}{2.401305(4)}       & \multicolumn{2}{c}{3.942667(3)}       & \multicolumn{2}{c}{9.057646(1)}       \\%
$a$ $sin\,i$ (R$_{\odot}$)                     & \multicolumn{2}{c}{21.452(95)}        & \multicolumn{2}{c}{14.343(203)}       & \multicolumn{2}{c}{24.899(117)}       & \multicolumn{2}{c}{53.655(469)}       \\%
$\gamma$ (kms$^{-1}$)                          & \multicolumn{2}{c}{174.8(2)}          & \multicolumn{2}{c}{153.4(3)}          & \multicolumn{2}{c}{152.3(2)}          & \multicolumn{2}{c}{139.1(8)}          \\%
K$_{1,2}$ (km s$^{-1}$)                        & 142.6(1.2)        & 324.7(1.3)        & 131.8(1.1)        & 170.5(1.3)        & 147.7(2.3)        & 174.5(2.8)        & 92.0(2.2)         & 207.9(2.4)        \\%
$e$                                            & \multicolumn{2}{c}{0.0}               & \multicolumn{2}{c}{0.0}               & \multicolumn{2}{c}{0.126(11)}         & \multicolumn{2}{c}{0.076(7)}          \\%
$\omega$(${\degr}$)                            & \multicolumn{2}{c}{90}                 & \multicolumn{2}{c}{90}                 & \multicolumn{2}{c}{252.3(2.4)}        & \multicolumn{2}{c}{58.4(3.1)}         \\%
$q$ (M$_2$/M$_1$)                              & \multicolumn{2}{c}{0.439(8)}          & \multicolumn{2}{c}{0.773(5)}          & \multicolumn{2}{c}{0.846(6)}          & \multicolumn{2}{c}{0.463(9)}          \\%
$i$ (${\degr}$)                                & \multicolumn{2}{c}{80.4(1)}           & \multicolumn{2}{c}{78.9(4)}           & \multicolumn{2}{c}{84.4(2)}           & \multicolumn{2}{c}{84.2(5)}           \\%
T$_{\rm eff\,1,2}$ (K)                         & 28\,510$^a$[fix]  & 18\,560(110)      & 12\,280$^a$[fix]  & 11\,990(140)      & 19\,420$^a$[fix]  & 17\,390(220)      & 32\,300$^a$[fix]  & 22\,820(150)      \\%
$\Omega_{1,2}$                                 & 2.7271(36)        & 2.7568            & 4.3289(41)        & 6.6640(44)        & 6.6976(23)        & 6.3021(54)        & 7.4728(39)        & 5.7796(52)        \\%
$A_{1,2}$                                      & \multicolumn{2}{c}{1.0[fix]}          & \multicolumn{2}{c}{1.0[fix]}          & \multicolumn{2}{c}{1.0[fix]}          & \multicolumn{2}{c}{1.0[fix]}          \\%
$g_{1,2}$                                      & \multicolumn{2}{c}{1.0[fix]}          & \multicolumn{2}{c}{1.0[fix]}          & \multicolumn{2}{c}{1.0[fix]}          & \multicolumn{2}{c}{1.0[fix]}          \\%
$X_{1,2}^c$                                    & 0.209             & 0.202             & 0.201             & 0.198             & 0.163             & 0.221             & 0.165             & 0.204             \\%
$l_{1}/(l_1 + l_2)$                            & \multicolumn{2}{c}{0.82(1)}           & \multicolumn{2}{c}{0.82(5)}           & \multicolumn{2}{c}{0.59(4)}           & \multicolumn{2}{c}{0.75(3)}           \\%
$r_{\rm 1,2}$                                  & 0.4535(2)         & 0.3021(2)         & 0.2846(2)         & 0.1398(2)         & 0.1751(6)         & 0.1694(8)         & 0.1438(2)         & 0.1048(3)         \\%
$\Sigma$ $W(O-C)^2$                            & \multicolumn{2}{c}{0.035}             & \multicolumn{2}{c}{0.031}             & \multicolumn{2}{c}{0.033}             & \multicolumn{2}{c}{0.096}             \\ 
\hline
Absolute parameters                            &&&&&&&&&&\\ 
\hline
M$_{1,2}$ (M$_{\odot}$)                        & 17.803(70)        & 7.813(71)         & 4.103(39)         & 3.172(44)         & 7.370(98)         & 6.238(96)         & 17.556(77)        & 8.128(79)         \\%
R$_{1,2}$ (R$_{\odot}$)                        & 9.866(42)         & 6.573(56)         & 4.160(55)         & 2.045(55)         & 4.389(95)         & 4.246(101)        & 7.755(52)         & 5.652(71)         \\%
$\log(g_{1,2})$ (cgs)                          & 3.70(3)           & 3.71(5)           & 3.81(5)           & 4.32(5)           & 3.98(8)           & 4.02(2)           & 3.90(6)           & 3.89(7)           \\%
$(v_{1,2}\sin i)_{\rm calc}$ (km s$^{-1}$)$^d$ & 214(6)            & 143(7)            & 88(4)             & 43(3)             & 56(4)             & 54(7)             & 43(4)             & 31(8)             \\%
$\log(L_{1,2}/L_{\odot})$                      & 4.852(71)         & 3.715(52)         & 2.553(48)         & 1.863(64)         & 3.451(37)         & 3.145(65)         & 4.774(31)         & 3.937(66)         \\%
M$_{\rm bol\,1,2}$ (mag)                       & -7.380(114)       & -4.537(115)       & -1.632(219)       & 0.091(310)        & -3.878(92)        & -3.113(163)       & -7.185(76)        & -5.091(163)       \\%
(m--M)$_V$ (mag)                               & \multicolumn{2}{c}{19.999(318)}       & \multicolumn{2}{c}{20.112(209)}       & \multicolumn{2}{c}{19.104(133)}       & \multicolumn{2}{c}{18.961(207)}       \\%
$E(B-V)$ (mag)                                 & \multicolumn{2}{c}{0.327(6)}          & \multicolumn{2}{c}{0.011(109)}        & \multicolumn{2}{c}{0.024(112)}        & \multicolumn{2}{c}{0.419(37)}         \\%
$d$ (kpc)$^b$                                  & \multicolumn{2}{c}{64.7(3.2)}         & \multicolumn{2}{c}{54.9(4.8)}         & \multicolumn{2}{c}{58.5(6.1)}         & \multicolumn{2}{c}{64.1(4.7)}         \\%
\hline
\end{tabular}     
\begin{minipage}{0.95\textwidth}
\footnotesize
\emph{Notes}: 
$^a$T$_{\rm eff\,_1}$ were found in atmospheric analysis in \S~\ref{sec:atm}.
$^b$The \jktabs\ distance is calculated from 2MASS magnitudes.
\end{minipage}
\end{table*}

\subsection{Star location within a cluster $vs.$ clustered star location}
\label{smc}
An unbiased evaluation enables us to examine the potential relationship between the NGC\,290 cluster 
and the stars referred to in our study. Eight spectroscopically analysed SMC eclipsing binaries 
are projected toward the open cluster NGC~290. In order to evaluate their possible physical association, we 
construct a consistent astrometric and structural reference framework for the cluster using \emph{Gaia}~DR3 
data. The cluster is treated as a reference environment rather than as the primary scientific objective. Structural 
parameters, mean kinematics, and fundamental astrophysical properties are derived to assess the membership status 
of the eight targets through combined spatial, kinematic, and astrophysical consistency tests. The SMC targets 
lie within the projected field of NGC\,290. To establish a reference cluster framework, we queried a circular 
region with $10'$ of radius centred on $(\mathrm{RA},\mathrm{Dec})=(12.801^\circ,-73.1617^\circ)$ \citep{2023A&A...672A.161M} 
from the \emph{Gaia}~DR3 archive, yielding 97,257 sources.

To reduce Milky Way foreground contamination, a parallax constraint of $\varpi \leq 0.2~\mathrm{mas}$ 
was applied \citep{2024A&A...681A..24S}. At the SMC distance of approximately $60~\mathrm{kpc}$, this threshold efficiently 
removes most nearby foreground stars. In addition, we adopted the completeness-oriented threshold (Tcut = 0.31) defined by 
\citet{2023A&A...672A..65J} to avoid excluding potential SMC members. This permissive threshold maximises completeness 
at the expense of a higher level of Milky Way contamination, which we account for in the subsequent structural and kinematic 
analysis. This reduced the sample to 12,162 high-probability SMC sources that remained within the $10'$ region. This filtered 
sample provides the SMC-dominated stellar population within which NGC\,290 is embedded.

To define the spatial extent of NGC\,290 and provide a reference for membership testing, a radial 
density profile (RDP) was constructed and fitted with a \citet{1962AJ.....67..471K} model using a minimum-$\chi^2$ 
approach. The resulting structural parameters indicate a core radius of $r_c = 0.280 \pm 0.046'$, and the limiting 
radius of $r_{\mathrm{lim}}=5'$. The corresponding concentration parameter, $C=\log(r_{\mathrm{lim}}/r_c)=1.4$, is 
consistent with previous determinations \citep{2021MNRAS.507.3312G} and indicates a moderately concentrated cluster 
with a compact core and a well-defined halo.

Cluster membership probabilities were computed in proper-motion space using the two-component mixture 
model of \citep{1998A&AS..133..387B}, which statistically separates cluster and field populations through a Gaussian 
description of their kinematic distributions. Parallaxes were used exclusively as a foreground filter and were not 
incorporated into the probability computation, because at the SMC distance the expected parallax ($\sim0.017~\mathrm{mas}$) 
is comparable to \emph{Gaia}~DR3 uncertainties and zero-point. Stars within $r_{\mathrm{lim}}$ were analysed, 
and those with $P\geq0.5$ were initially classified as probable members. Residual contamination in proper-motion space 
was further reduced by inspecting the colour-magnitude diagram (CMD) morphology and the radial density profile (RDP) behaviour.

To derive robust mean kinematic properties, only stars within $r \leq 3r_c \approx 1'$ were considered, 
corresponding to the highest-density central region. From 74 probable members in this inner zone, the mean proper motion 
components were determined as $(\mu_{\alpha}\cos\delta,\mu_{\delta})=(0.609 \pm 0.169,-1.303 \pm 0.172)$ mas~yr$^{-1}$, 
consistent with the SMC cluster kinematics reported by \citet{2023A&A...672A.161M}. One inner-region member has a 
{\it Gaia} DR3 radial velocity measurement ($V_r = 170.74 \pm 1.27$ km~s$^{-1}$), adopted here as a representative 
cluster value.


Fundamental cluster parameters were derived by fitting the non-rotating PARSEC isochrones of 
\citet{2012MNRAS.427..127B} to the (G) versus $(BP-RP)$ CMD constructed from stars within $r\leq1'$. The 
best-fitting solution yields an age of $20 \pm 5~\mathrm{Myr}$ ($\log t = 7.3$), a distance of 
$63.4 \pm 2.8~\mathrm{kpc}$ corresponding to a distance modulus $\mu_G = 19.363 \pm 0.096~\mathrm{mag}$, 
reddening values of $E(BP-RP)=0.188 \pm 0.05~\mathrm{mag}$ and $E(B-V)=0.133 \pm 0.035~\mathrm{mag}$, and 
a metallicity of $[\mathrm{Fe/H}] = -0.7~\mathrm{dex}$. These values fall within the range of previous 
determinations \citep[e.g.,][]{2021A&A...647A.135N,2023A&A...672A.161M} and are adopted solely to provide 
a homogeneous reference for comparison with the target systems.

The eight SMC targets were evaluated for possible association with NGC\,290 through combined 
spatial, kinematic, and astrophysical consistency tests. Seven of the stars lie within the adopted limiting 
radius of $5'$, whereas OGLE\,SMC-ECL-1105 is located at a projected distance of $19'$ from the cluster centre. Although 
OGLE\,SMC-ECL-1105 exhibits proper-motion components consistent with the cluster mean within $1\sigma$ uncertainties; its 
large angular separation places it well beyond the region where the cluster density significantly exceeds the background 
level. Its spatial location, therefore, makes a bound association with NGC\,290 unlikely.

\begin{figure*}
\centering
\includegraphics[width=\linewidth]{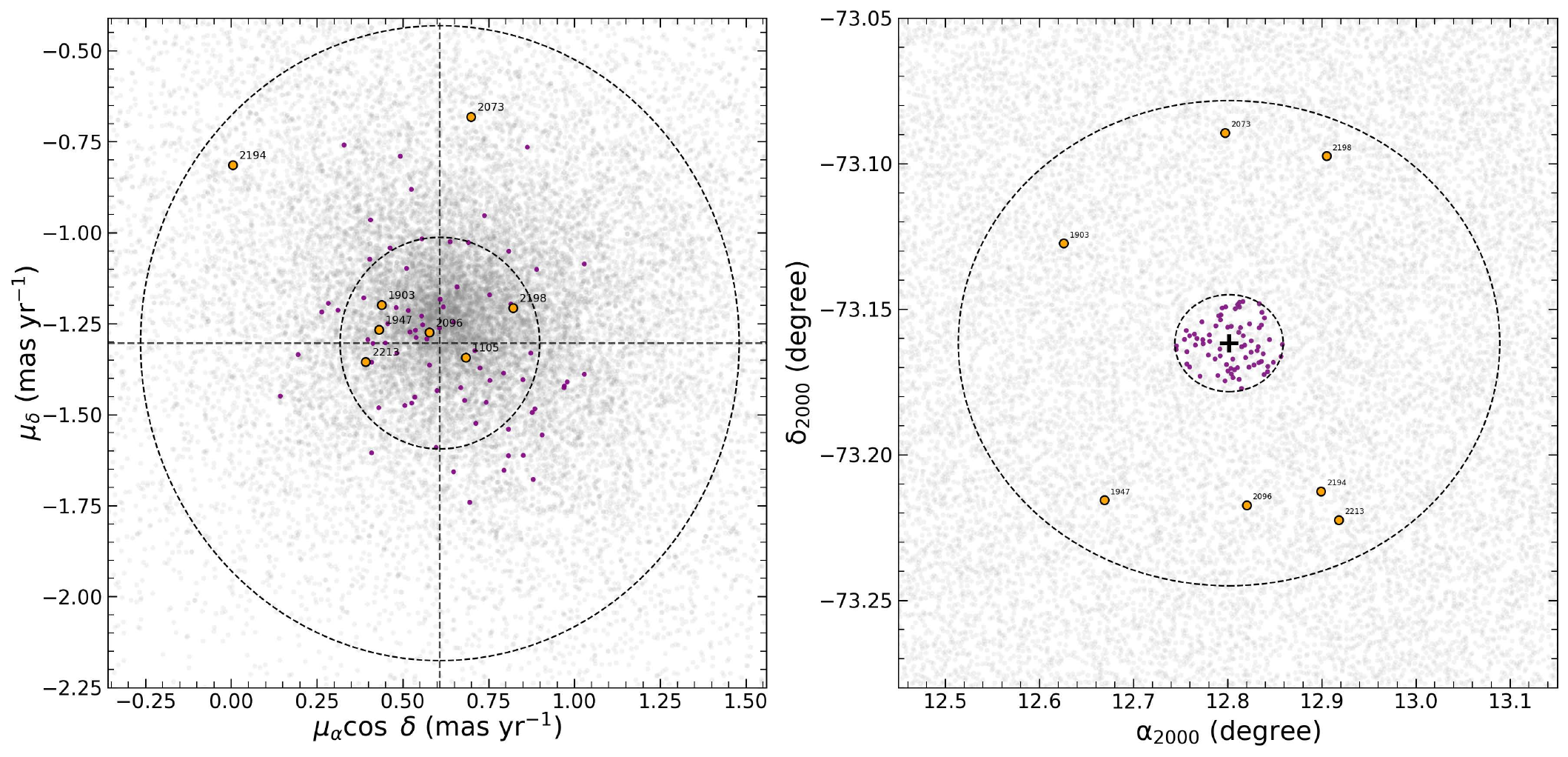}
\caption{Left: vector point diagram of stars projected toward NGC~290. Small dots denote probable cluster members 
within the central region ($r \leq 1'$), shaded symbols represent field stars, and filled circles mark the eight SMC 
binaries. Dashed circles indicate the $1\sigma$ and $3\sigma$ boundaries of the cluster proper-motion distribution. Right: 
spatial distribution of stars within a $10'$ region. Dashed circles mark the inner region ($r \leq 1'$) and the limiting 
radius ($r_{\rm lim}=5'$). SMC-1105, located at $19'$ from the centre, lies outside the plotted spatial field.}
\label{fig:vpd}
\end{figure*}

For each star, the deviation of its proper motion from the cluster mean was quantified using an 
uncertainty-weighted metric. This analysis indicates that OGLE\,SMC-ECL-2194 is best interpreted as a field star, while 
OGLE\,SMC-ECL-2073 occupies a borderline position at the $2$--$3\sigma$ level. The remaining stars are consistent with 
cluster membership in proper-motion space. A secondary probability computation using the \citet{1998A&AS..133..387B} 
formalism confirms that OGLE\,SMC-ECL-2194 and OGLE\,SMC-ECL-2073 have probabilities close to zero, whereas the other 
six stars satisfy $P>0.5$. Figure~\ref{fig:vpd} illustrates the proper-motion distribution and the spatial configuration 
of the cluster members and the eight targets.

\begin{figure*}
\centering
\includegraphics[width=0.6\linewidth]{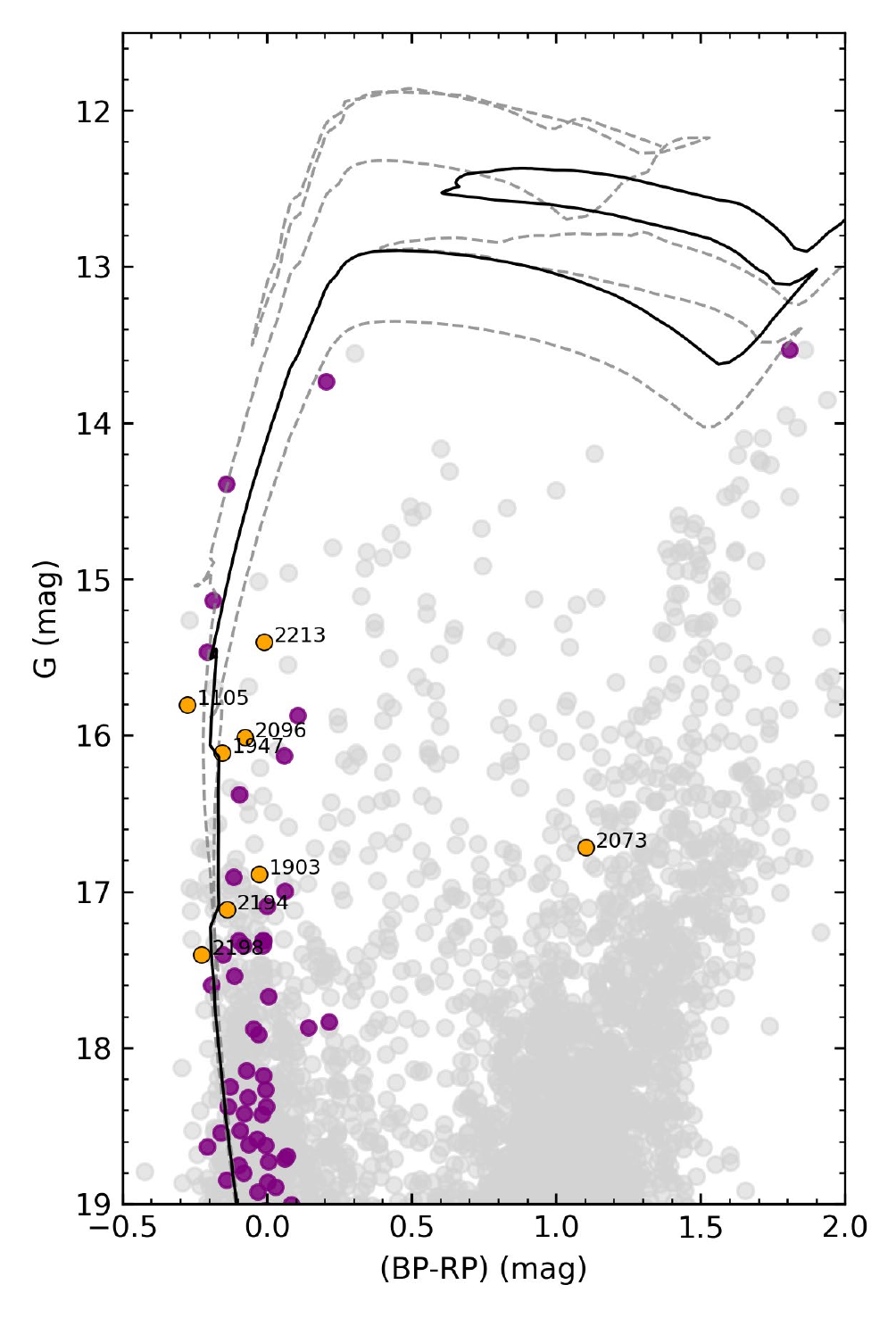}
\caption{Colour--magnitude diagram of NGC~290 for stars within $r \leq 1'$. Solid circle markers denote probable cluster 
members, and shaded markers represent other projected stars. The solid curve shows the best-fitting PARSEC isochrone 
corresponding to $\log t = 7.3$, $d = 63.4$ kpc, $E(B-V)=0.13$ mag, and $[\mathrm{Fe/H}]=-0.7$\,dex. The eight SMC 
eclipsing binaries are shown as filled circles.}
\label{fig:cmd}
\end{figure*}

The final assessment incorporates stellar ages, distances, and radial velocities derived from the 
spectroscopic analysis. Figure~\ref{fig:cmd} presents the colour--magnitude diagram of NGC\,290, including the probable 
cluster members, the best-fitting PARSEC isochrone, and the eight SMC targets. In this comparison, OGLE\,SMC-ECL-1903 
appears as the most plausible candidate for association with NGC~290 based on its projected proximity and its location 
relative to the cluster sequence; however, the available constraints are insufficient for a secure membership assignment. The 
radial velocity of OGLE\,SMC-ECL-2213 is consistent with the provisional cluster value, although its age appears younger 
than that of NGC\,290. Within this framework, the cluster properties provide the environmental context for interpreting 
the physical nature of the eight spectroscopically analysed SMC stars.


Building on the global spatial distribution discussed above, we note that sky-plane proximity 
alone does not provide a reliable basis for membership without a homogeneous structural and astrometric characterisation 
of the cluster. The \emph{Gaia}~DR3-based framework established here enables a quantitative evaluation of membership 
likelihood. Using the derived constraints on distance, proper motion, and radial velocity, we additionally examine 
the present-day dynamical state of each target within the Magellanic gravitational potential under two scenarios: an 
isolated SMC potential and a combined SMC+LMC configuration. This two-tiered approach places the concept of stellar 
``isolation'' within a coherent environmental and dynamical context.

Kinematic analysis provides the primary basis for evaluating the likelihood of cluster association, while 
orbit integrations offer complementary context on the present-day dynamical state of the systems. Relying solely on static 
indicators, such as projected proximity and instantaneous proper motion similarities, is insufficient to confirm a common 
origin. The phase-space structure in this region may be influenced by the irregular morphology of the SMC, uncertainties 
in distances and velocities, and the tidal influence of the Large Magellanic Cloud (LMC). Therefore, to model the mass 
distribution in the inner regions where star clusters predominantly form, we adopt the analytical \texttt{Hernquist} 
profile \citep{1990ApJ...356..359H}. Orbital analyses were performed using the \texttt{galpy}\footnote{\url{https://github.com/jobovy/galpy}} \citep{2015ApJS..216...29B} library to integrate the 
orbits of the NGC\,290 cluster and the eight SMC stars under a \texttt{Hernquist} potential in two scenarios: (\textit{i}) the 
gravitational potential of the SMC alone, and (\textit{ii}) the combined total potential of the SMC and LMC.

In these analyses, the orbits of the NGC 290 cluster and each of the eight investigated stars were 
modelled starting from their observed input parameters ($\alpha, \delta, d, \mu_{\alpha} \cos \delta, \mu_\delta, V_r$) to test 
for a shared dynamical origin. Although the \texttt{Hernquist} profile assumes spherical symmetry and thus cannot fully 
capture the tidally distorted morphology of the SMC \citep{2009MNRAS.395..342B}, it provides a first-order approximation 
to the enclosed distribution in the inner few kpc. Accordingly, in our simulations, we adopted a mass of 
$M_{\rm SMC} = 2.4 \times 10^9 M_\odot$, corresponding to the enclosed within a galactocentric radius of 
$r \leq 3\,\mathrm{kpc}$, as derived from the H\,I rotation curve \citep{2004ApJ...604..176S}, and a scale radius of 
$a_{\rm SMC} = 2.0$\,kpc \citep{2005MNRAS.356..680B, 2012MNRAS.421.2109B}. In the combined scenario, the gravitational 
potential of each galactic component was treated as an independent \texttt{Hernquist} sphere. For the LMC potential, we 
adopted a mass of $M_{\rm LMC} = 1.8 \times 10^{11} M_\odot$ and a scale radius of $a_{\rm LMC} = 20.0$\,kpc, consistent 
with the total mass range constrained by stellar stream perturbations \citep{2019MNRAS.487.2685E,2021ApJ...923..149S} and 
satellite kinematics \citep{2024MNRAS.527..437V}. Within this SMC+LMC framework, the LMC was positioned as a static external 
potential at its present-day location; while a simplification, this approach provides a robust and sufficient basis for 
evaluating the instantaneous boundness of each target.

\begin{figure*}
\includegraphics[width=0.49\textwidth]{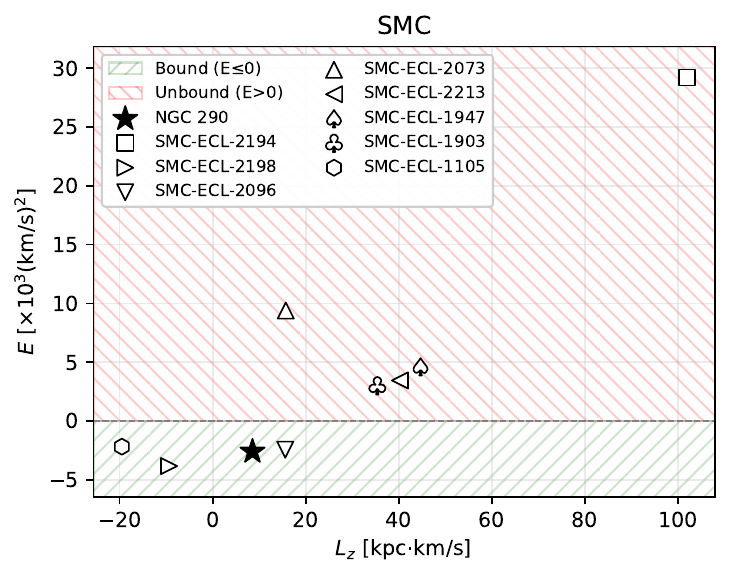}
\includegraphics[width=0.49\textwidth]{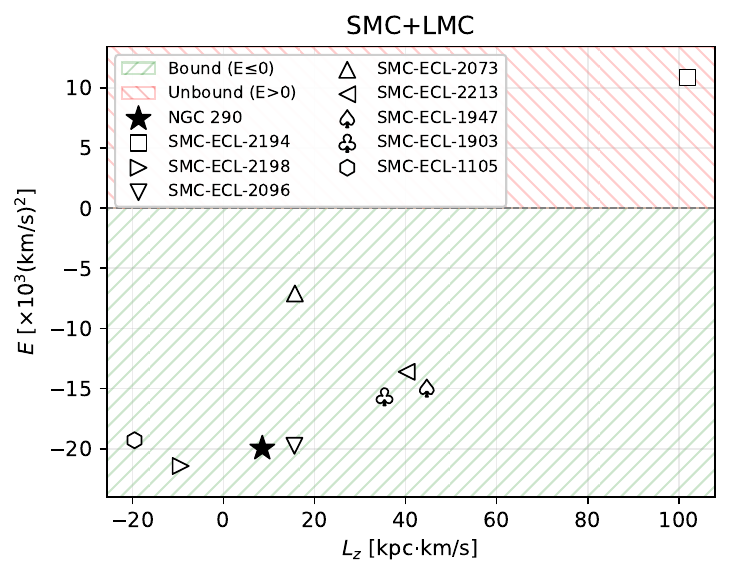}
\caption{The Lindblad diagrams under the SMC potential (left) and the combined SMC+LMC potentials (right). 
The dashed areas represent the bound and unbound regions (\textit{see text for more information}).
%
}
\label{lindbalt}
\end{figure*}

The orbital motions of star clusters and individual stars within a galactic potential provide critical 
insights into the dynamical structure and evolutionary history of the system. Diagnostic diagrams based on the conserved 
quantities of total orbital energy ($E$) and the vertical component of angular momentum ($L_z$), commonly referred to as 
Lindblad diagrams, provide a standard tool for kinematic classification \citep{2008gady.book.....B}. In this normalisation, the 
condition $E > 0$ corresponds to a formally unbound state. To evaluate the boundness of the targets, we first analysed the 
kinematics under the isolated SMC potential (Scenario~\textit{i}). The resulting $E$ versus $L_z$ diagram (Figure\,\ref{lindbalt}) 
reveals a clear kinematic bifurcation.

%
%
%
%

As shown in Figure\,\ref{lindbalt}, the NGC\,290 cluster and three SMC targets are gravitationally 
bound to the SMC ($E < 0$). Specifically, NGC\,290 is deeply embedded within the potential well, locating at 
$E\approx-2500$ (km/s)$^2$ and $L_z \approx 10$ kpc km/s, which indicates a stable orbit consistent with the 
dynamical centre of the SMC. Among the bound stars, SMC-ECL-2198 and SMC-ECL-2096 exhibit energy levels comparable 
to NGC\,290, whereas SMC-ECL-1105 displays a retrograde or highly inclined orbit with negative angular momentum 
($L_z \approx -20$ kpc km/s) while remaining bound. Conversely, the remaining systems occupy the unbound region 
($E > 0$). Most notably, SMC-ECL-2194 emerges as an extreme outlier with $E \approx 29000$ (km/s)$^2$ and 
$L_z \approx 100$ kpc km/s. SMC-ECL-2073 is also distinctly unbound ($E \approx 9000$ (km/s)$^2$), while the 
others cluster in the range of $E\approx3000-5000$ (km/s)$^2$.

However, the interaction between the local SMC potential and the more massive neighbouring LMC can 
radically alter the orbital energies of these systems \citep{2012MNRAS.421.2109B}. Evaluating the dynamical states 
under the combined SMC+LMC potential (Scenario \textit{ii}, Figure\,\ref{lindbalt}) demonstrates a significant downward vertical 
shift in the energy distribution.

The inclusion of the massive LMC deepens the total gravitational pool \citep{2005MNRAS.356..680B}, causing 
the majority of the stellar population that appeared unbound in the isolated SMC model to drop into the $E < 0$ regime. For 
instance, the energy of SMC-ECL-2073 decreases drastically from $+9000$ to approximately $-7000$ (km/s)$^2$, illustrating 
that even if an object escapes the immediate vicinity of the SMC, it remains trapped by the LMC's deep gravitational 
well. Similarly, the natively bound NGC\,290 cluster becomes even more tightly bound to the system, shifting to 
$E \approx -20000$ (km/s)$^2$.

Despite this systematic energy reduction, the extreme outlier SMC-ECL-2194 remains formally unbound, with 
its energy decreasing only to $E \approx 11000$ (km/s)$^2$. This behaviour is consistent with a high-velocity runaway 
ejected via a supernova in a binary system (the `binary supernova scenario'; \citealt{1961BAN....15..265B}) or through 
dynamical interactions in dense stellar environments \citep{1967BOTT....4...86P}. Alternatively, it may represent a 
high-velocity interloper traversing the region, potentially linked to three-body interactions involving a massive black 
hole or dense cluster cores \citep{1988Natur.331..687H, 2015ARA&A..53...15B}. Overall, this comparison highlights that 
kinematic interpretation in this region should not rely solely on an isolated SMC potential. Systems that appear to be 
escaping the SMC may remain bound to the broader Magellanic gravitational environment and could trace large-scale 
structures such as the Magellanic Bridge \citep{1996MNRAS.278..191G,2007ApJ...668..949B}.

\begin{figure*}
\centering	
\includegraphics[width=0.32\textwidth]{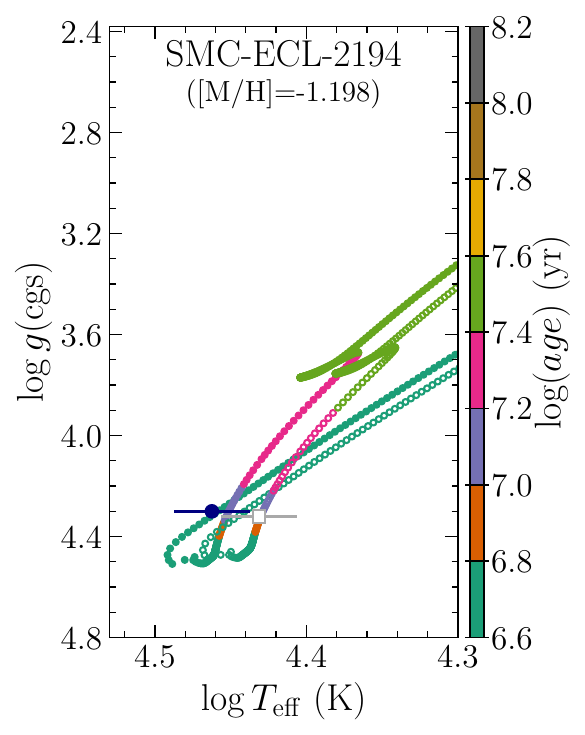}
\includegraphics[width=0.32\textwidth]{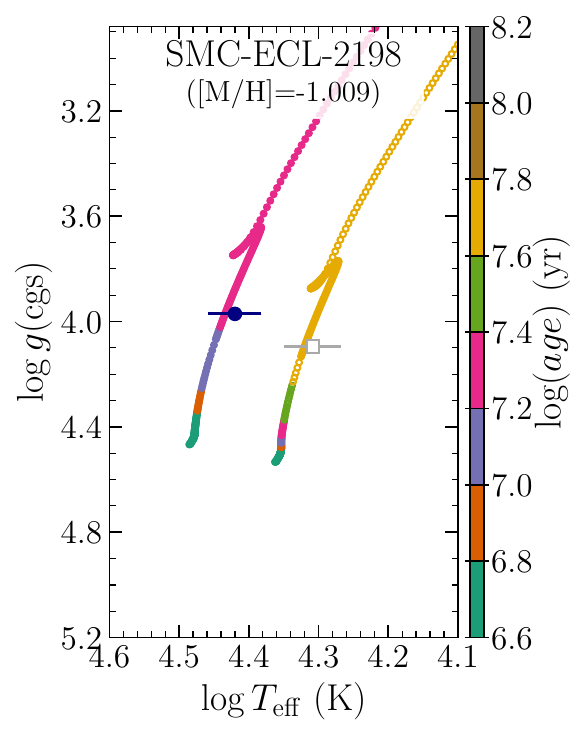}
\includegraphics[width=0.32\textwidth]{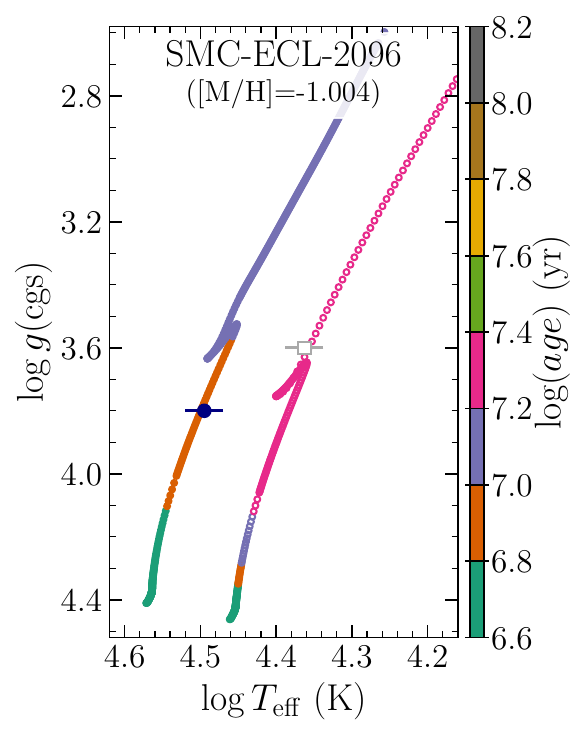}
\includegraphics[width=0.32\textwidth]{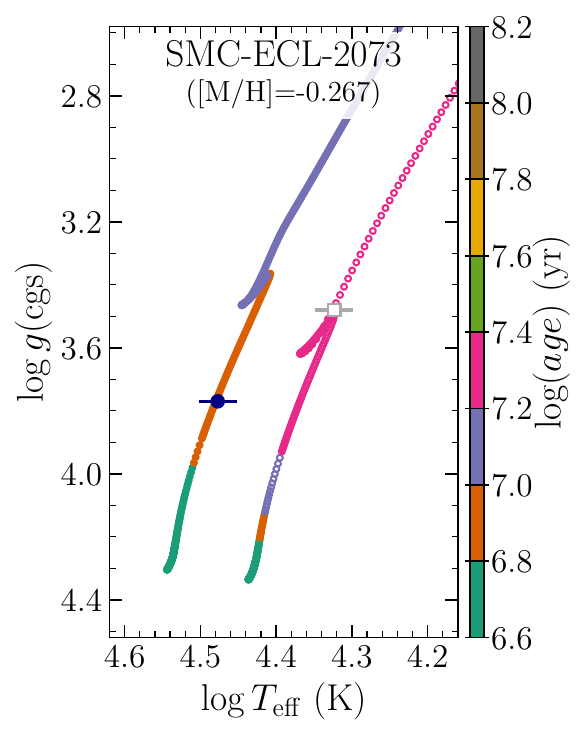}
\includegraphics[width=0.32\textwidth]{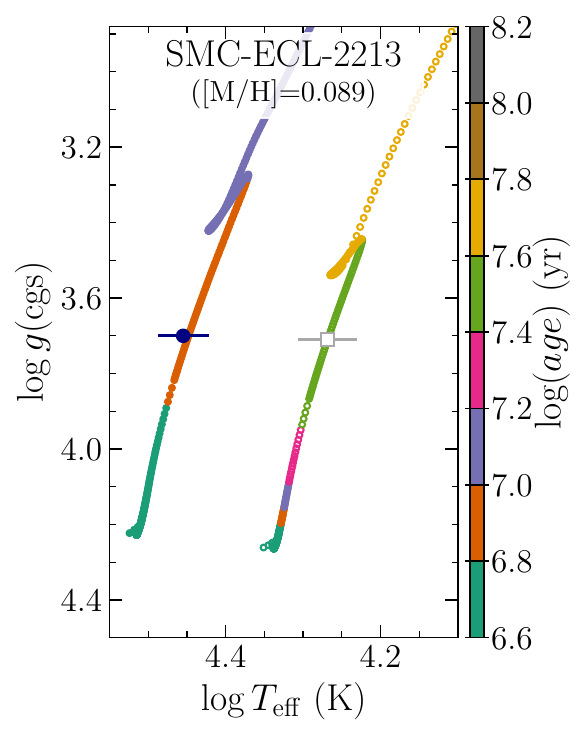}
\includegraphics[width=0.32\textwidth]{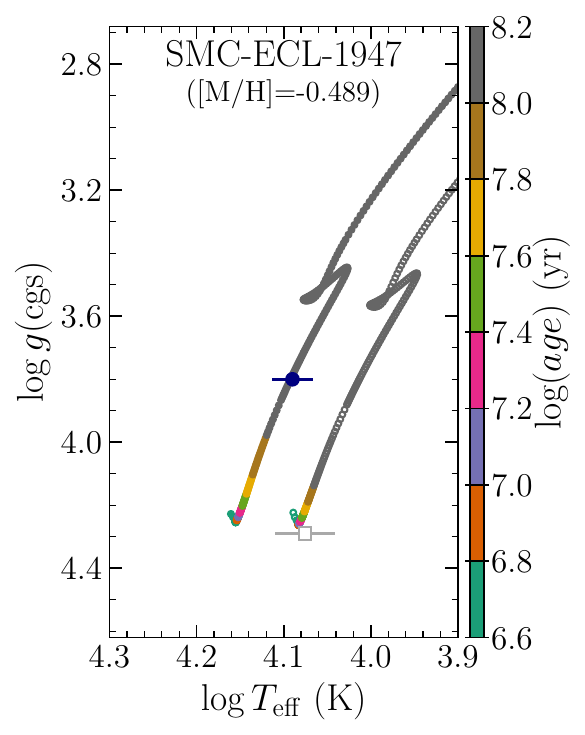}
\includegraphics[width=0.32\textwidth]{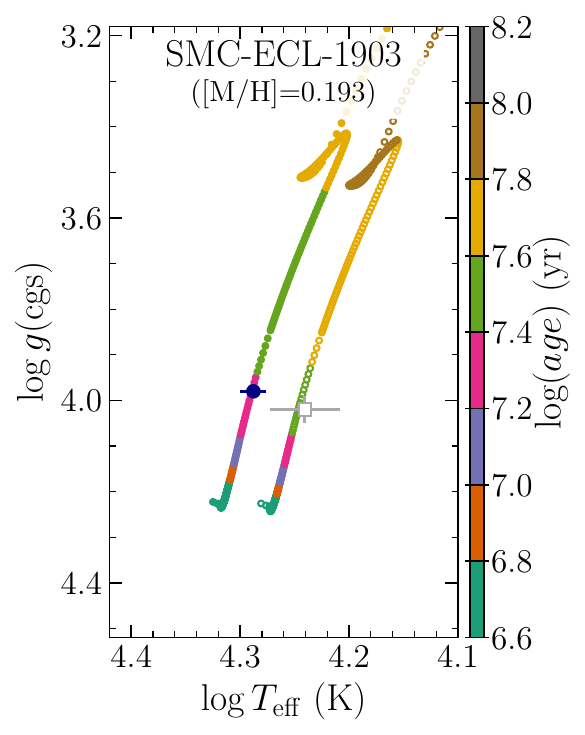}
\includegraphics[width=0.32\textwidth]{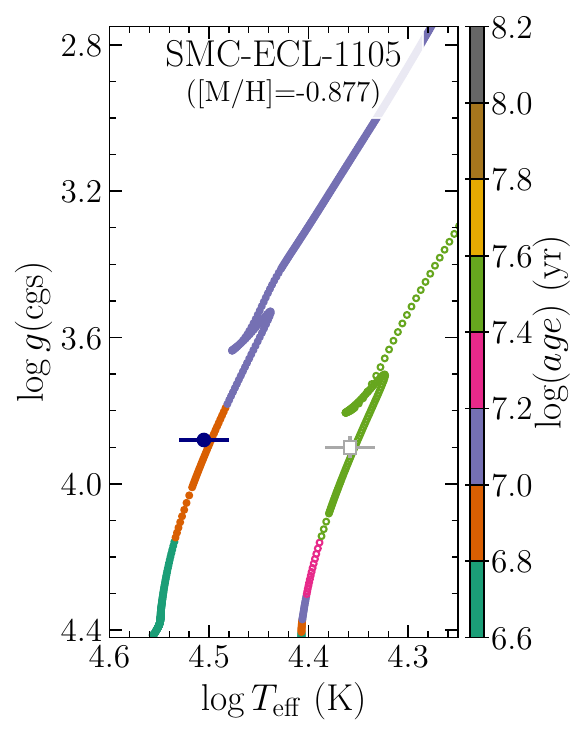}
    \caption{Our findings are compared with the MESA evolution tracks on \emph{Kiel} diagrams within the 
$\log T_{\rm eff}$ / $\log(g)$ planes. The primary component is shown by filled circles, whereas the 
secondary components are shown as open squares, each with error bars as outlined in Tables\,\ref{salt_table1} 
and \ref{salt_table2}. The coloured points represent the evolutionary tracks linked to the calculated 
component masses, with the colours denoting stellar age, as indicated by the colour bars on the right 
side of the plots. }
\label{evo}
\end{figure*}

In the following section, we examine whether the ages inferred from the binary solutions 
are consistent with the cluster reference framework, and use this comparison to refine the interpretation 
of cluster association versus field origin.

\section{Evolutionary status}
\label{sec:evo}
A comparison of the components of each system with the theoretical evolution tracks in the \textit{Kiel} diagram 
($\log T_{\rm eff}$\,-\,$\log(g)$ diagram) can be made by merging the results of the spectroscopic and photometric analyses. 
This process leads to an estimation of the evolutionary status of each component.
The main benefit of the \textit{Kiel} diagram is the ability to simplify the comparison of stars with 
predictions of stellar evolution, all without requiring any prior knowledge of their distance. An exemplary 
application of this technique can be found in \citet{1976A&A....52...11K}. The relatively large uncertainties of spectroscopic gravity determinations often negatively 
impact the comparison with stellar evolution calculations. Thus, the $\log g$ value of each star in the \emph{Kiel} diagram 
has been determined by the results of its combined light and velocity curve analysis, rather than by atmospheric 
parameters, due to conflicting factors. 
By combining the stellar abundances estimated from the atmospheric analyses, we successfully estimated the stellar ages from the evolutionary tracks.

To show the evolutionary status of our samples, we use the \textsc{mesa} Isochrones and Stellar Tracks 
\citep[MIST\footnote{\url{https://mist.science/}} v1.2,][]{2016ApJS..222....8D,2016ApJ...823..102C}, which 
is generated by the Modules for Experiments in Stellar Astrophysics (\textsc{mesa\footnote{\url{https://mesastar.org/}}}) 
package \citep{2011ApJS..192....3P,2013ApJS..208....4P,2015ApJS..220...15P} and serves as a thorough and updated 
version of the stellar evolution model for a wide range of metallicities. The evolutionary tracks shown in Figure\,\ref{evo} 
were generated using the MIST web interpolator, which made the task easier. Then we compared our results with these 
evolutionary tracks.

The \textit{Kiel} diagrams of SMC-ECL-2194 and SMC-ECL-2198 represent a complex case. The light and radial velocity 
curves of SMC-ECL-2194 provide reliable results with small errors; however, they do not fully match the predicted 
evolutionary paths of the system's components. However, when plotted on the \emph{Kiel} diagram, utilising 
temperature and $\log (g)$ values derived from atmospheric analysis, SMC-ECL-2194 demonstrates increased 
consistency. The circumstances surrounding SMC-ECL-2198 are totally separate. The atmospheric analysis of 
SMC-ECL-2198 reveals values that, when plotted on the HR diagram, demonstrate that the system's components 
adhere to their expected evolutionary path. Nevertheless, its position on the HR diagram, as indicated by 
the light curve analysis results, contradicts established evolutionary trends. The discrepancies in the 
evolutionary paths of both systems, SMC-ECL-2194 and SMC-ECL-2198, are proposed to arise from the light 
curves, which exhibit limited sensitivity and a restricted number of data points. As visible in 
Figure\,\ref{fig_LC_RV_plot1}, the significant scattering of the light curve of the star SMC-ECL-2198 
significantly affects the reliability of the analysis results, which may be attributed to incorrect 
modelling. We would like to highlight that the results of the solution cannot be considered reliable 
with respect to the absolute parameters of the secondary star. However, the atmospheric analyses utilising 
GIRAFFE data, despite its low spectral resolution, are generally reliable for all stars. Observations 
of additional stars provide evidence for this hypothesis. The considerable scattering and restricted data 
in the light curves of these two systems may inherently explain the discrepancies with the evolutionary 
pathways.

A comparison of the astrophysical parameters of the SMC-ECL-2096 and SMC-ECL-2073 systems, derived from 
spectroscopic and combined photometric+radial velocity analyses, shows minor discrepancies as well as 
notable similarities. The analysis of these binary systems represents significant advances compared to 
earlier research. The companion stars in both systems have filled their Roche lobes, and the two secondaries 
are located significantly away from the zero-age main sequence. The systems show notable similarities 
regarding essential astrophysical characteristics. This feature allows for the identification of different 
stellar parameters that may assist in clarifying the evolutionary status of the systems. This sophisticated 
system is especially intriguing, as it provides observational constraints for theoretical binary evolutionary 
models. Similar analyses have been performed for B-type stars \citep{2009MNRAS.400..791P}, but investigations 
involving systems made up of two evolved early-type stars are quite uncommon \citep[e.g., Plaskett's star as 
mentioned by][]{2008A&A...489..713L}. Analysis of these two systems emphasises the need to develop theoretical 
binary evolutionary models. 

Overall, the majority of the analysed systems are unlikely to be members of a bound stellar 
cluster in the SMC. Nevertheless, one or two cases remain plausible within the current astrometric and evolutionary 
uncertainties, and therefore cannot be excluded on firm grounds.

The systems referred to as field stars can be formed of isolated stars, recognised for their existence, as there
are isolated massive stars that do not have any nearby massive companions associated with them. Studies examining 
observations \citep{2002ApJS..141...81M} and statistical analyses \citep{2004MNRAS.348..187W} indicate that the 
initial mass function of field stars differs from that focused in stellar clusters. Additionally, theories propose 
that various physical factors may influence the formation of different massive stars in the field 
\citep[e.g.,][]{2003ApJ...592..975L}. The fundamental question for theories regarding massive star formation 
is whether OB stars are capable of forming in isolation. The inquiry into star formation, specifically regarding 
the potential for massive stars to evolve in isolation versus within stellar clusters, holds significant relevance 
for the understanding of massive star formation theories and the stellar initial mass function within entire 
galaxies. Although this question seems straightforward, providing an answer is quite complex. The study of these 
eight stars serves as an exemplary case of isolated Galactic massive star formation, primarily due to the presence 
of nearby indicators of recent star formation in the SMC, so that the more precise estimate of the absolute 
parameters is crucial.

\section{Summary and discussion}
\label{sec:sum_disc}
The stellar population of the observable SMC, which represents its history of interaction with the 
LMC, is what defines it. The connection between their distances, metallicities, and ages helps to explain 
these effects. We selected eight stars that we believe exemplify this representation and have not been 
previously studied. We derived the fundamental physical quantities, such as masses and radii, for all 
our systems with moderate precision and ooked for isolated stars in a galaxy outside of the Milky Way's 
immediate vicinity.

All eight stars are classified as massive stars and share similarities in terms of temperature, radius, and mass, with 
the exception of SMC-ECL-1947 and SMC-ECL-1903. While all analyses of the two systems were carried out efficiently 
and their astrophysical parameters were obtained with relatively small errors, there is a lack of agreement with the 
theoretical evolutionary paths. The systems SMC-ECL-2194 and SMC-ECL-2198 symbolise the primary challenge for obtaining 
an easy understanding of the discrepancies between the observed and predicted positions of the objects in the 
Hertzsprung-Russell diagram. The observed spectra's low signal-to-noise ratio (S/N) restricts the precision of the 
derived surface temperatures and metallicities from spectroscopy. We achieved temperature uncertainties of less than 
300\,K for six stars, while the other two stars showed uncertainties of approximately 700\,K. This section of the 
analysis suggests that there are likely still opportunities to enhance these measurements through further spectroscopic 
observations. 

The PLATO satellite presents a substantial opportunity to deepen our understanding of these sources. Upon 
becoming operational, the PLATO telescope's field of view will include approximately fifty per cent of the 
SMC (Southworth et al. 2026, \textit{in prep.}). We believe that this will provide an opportunity to 
enhance our models and validate the concept that the targets are intrinsically variable.

A further important aspect of this study concerns the environmental context of the eight 
systems. Although the targets are projected toward the open cluster NGC\,290, a homogeneous \emph{Gaia}~DR3-based
structural and kinematic analysis demonstrates that firm cluster membership can only be established in a limited 
number of cases. Seven systems lie within the adopted limiting radius of the cluster, yet proper-motion and spatial 
diagnostics indicate that at least two objects (SMC-ECL-2194 and SMC-ECL-2073) are more consistent with a field 
classification. OGLE\,SMC-ECL-1903 emerges as the most plausible candidate for association with NGC\,290; however, the 
available astrometric and evolutionary constraints remain insufficient for a definitive membership assessment. We 
note that probabilistic membership assignments based on proper-motion consistency do not automatically guarantee 
long-term dynamical association.

The orbital analysis further shows that several systems that appear formally unbound in an isolated 
SMC potential become bound when the gravitational influence of the LMC is included. This highlights the importance of 
adopting a global Magellanic framework when interpreting stellar kinematics in this region. In particular, SMC-ECL-2194 
remains dynamically distinct even in the combined SMC+LMC configuration, suggesting a possible high-velocity or runaway 
nature. Overall, our results indicate that the majority of the systems are not securely associated with NGC\,290, but are 
instead embedded within the broader dynamical structure of the Magellanic System.

These findings imply that apparent isolation in projection does not necessarily correspond to dynamical 
isolation. The combination of precise astrometry and orbit integration demonstrates that the concept of “isolated massive 
stars” in the SMC must be evaluated within the full Magellanic gravitational context rather than solely through sky-plane 
proximity.

\section{acknowledgements}
We thank the anonymous reviewer for their careful reading of our
manuscript and their many insightful comments and suggestions.
The following internet-based resources were used in research for this paper: the NASA 
Astrophysics Data System; the SIMBAD database operated at CDS; as well as the open-source Python packages 
ASTROPY {\url{http://www.astropy.org}} \citep{2013A&A...558A..33A}, MATPLOTLIB \citep{2007CSE.....9...90H}, NUMPY \citep{2020Natur.585..357H}, 
and PANDAS \citep{2020zndo...3630805R}. This work is based on data from the \kep\ mission. \kep\ was 
competitively selected as the tenth Discovery mission. Funding for this mission is provided by NASA's 
Science Mission Directorate. The photometric data were obtained from the Mikulski Archive for Space 
Telescopes (MAST). This work has also made use of data from the European Space Agency (ESA) mission 
\gaia\ (\url{http://www.cosmos.esa.int/gaia}), processed by the \gaia\ Data Processing and Analysis 
Consortium (DPAC, \url{http://www.cosmos.esa.int/web/gaia/dpac/consortium}). This research has also 
made use of the "Aladin sky atlas" developed at CDS, Strasbourg Observatory, France. The used spectra 
in this study were obtained by using the available spectra that were collected by the program IDs; 
60.A-9022(C), 0104.A-9001(A), 0106.A-9006(A), 073.C-0337(A), and 178.D-0361(F). This article is a 
part of the PhD thesis of MK.
%


\bibliography{Bibliography}{}
\bibliographystyle{aasjournalv7}

\clearpage
\appendix
\fontsize{9}{10}\selectfont
\begin{center}

\captionof{table}{Measured radial velocities (RV) of the components in each target system, along with statistical 
uncertainties ($\sigma_{RV}$). The signal-to-noise ratio (SNR) of each observation is shown in the last 
column.}\label{table_radial_velocities}
\end{center}
\begin{center}
\resizebox{15.0cm}{!}{%
\begin{tabular}[t]{cccccc}
\hline
HJD           & RV$_{1}$ & $\sigma_{RV_{1}}$ & RV$_{2}$ & $\sigma_{RV_{2}}$ & \multirow{2}{*}{Instrument} \\
(24\,00000+)  & (\kms)   & (\kms)            & (\kms)   & (\kms)            &                             \\
\hline\noalign{\smallskip}
\multicolumn{6}{l}{SMC-ECL-2194} \\%
\hline
52959.54228   & -47.7    & 6.3               & 378.7    & 7.3               & GIRAFFE                     \\
52959.71149   & -78.0    & 3.1               & 417.6    & 8.3               & GIRAFFE                     \\
52960.53937   & 259.9    & 5.1               &  51.4    & 3.6               & GIRAFFE                     \\
52960.67779   & 111.7    & 2.3               & 216.9    & 3.5               & GIRAFFE                     \\
52961.53366   & 386.2    & 3.7               & -86.1    & 0.1               & GIRAFFE                     \\
52961.70896   & 361.2    & 5.7               & -61.0    & 2.3               & GIRAFFE                     \\
52962.54352   & 113.7    & 3.7               & 214.1    & 4.3               & GIRAFFE                     \\
52962.72694   & 323.4    & 5.6               &  -4.7    & 3.5               & GIRAFFE                     \\
52963.53348   & -78.7    & 1.4               & 417.2    & 3.9               & GIRAFFE                     \\
52964.53579   & 170.1    & 2.4               & 153.2    & 2.9               & GIRAFFE                     \\
52964.55598   & 148.2    & 1.4               & 175.2    & 1.9               & GIRAFFE                     \\
52964.75008   & -36.7    & 4.1               & 376.2    & 1.9               & GIRAFFE                     \\
52965.52868   & 399.5    & 0.2               & -95.7    & 1.9               & GIRAFFE                     \\
52965.70366   & 309.2    & 3.3               &   2.7    & 1.9               & GIRAFFE                     \\
52966.54541   & 222.7    & 3.5               &  92.2    & 1.9               & GIRAFFE                     \\
52966.70023   & 366.6    & 4.2               & -63.7    & 1.9               & GIRAFFE                     \\
56567.64033   & -37.0    & 2.2               & 370.2    & 1.9               & GIRAFFE                     \\
56569.60045   & 336.5    & 2.5               & -29.7    & 1.9               & GIRAFFE                     \\
56587.53031   & 333.2    & 1.6               & -25.7    & 1.9               & GIRAFFE                     \\
56599.72156   & 190.2    & 1.7               & 124.7    & 1.9               & GIRAFFE                     \\
56621.67172   & 365.0    & 1.7               & -67.7    & 1.9               & GIRAFFE                     \\
\hline\noalign{\smallskip}
\multicolumn{6}{l}{SMC-ECL-2198} \\
\hline
52959.54229   & 364.1    & 1.3               & -146.8   & 5.3               & GIRAFFE                     \\
52959.71150   & 289.1    & 1.3               &  -45.1   & 5.3               & GIRAFFE                     \\
52960.53938   & 162.3    & 1.3               &  124.1   & 5.3               & GIRAFFE                     \\
52960.67780   & 288.1    & 1.3               &  -51.1   & 5.3               & GIRAFFE                     \\
52961.53366   & -75.1    & 1.3               &  433.1   & 5.3               & GIRAFFE                     \\
52961.70897   &   1.1    & 1.3               &  339.1   & 5.3               & GIRAFFE                     \\
52962.54352   & 122.1    & 1.3               &  180.1   & 5.3               & GIRAFFE                     \\
52963.53348   & 365.1    & 1.3               & -146.8   & 5.3               & GIRAFFE                     \\
52964.53580   & 168.5    & 1.3               &  117.1   & 5.3               & GIRAFFE                     \\
52964.55599   & 185.1    & 1.3               &   88.1   & 5.3               & GIRAFFE                     \\
52964.75009   & 340.1    & 1.3               & -118.8   & 5.3               & GIRAFFE                     \\
52965.52869   & -75.5    & 1.3               &  435.1   & 5.3               & GIRAFFE                     \\
52965.70367   &  14.1    & 1.3               &  324.1   & 5.3               & GIRAFFE                     \\
52966.54542   &  98.1    & 1.3               &  202.1   & 5.3               & GIRAFFE                     \\
52966.70024   & -29.1    & 1.3               &  372.1   & 5.3               & GIRAFFE                     \\
56621.67173   & 205.1    & 1.3               &   69.3   & 5.3               & GIRAFFE                     \\
\hline\noalign{\smallskip}
\multicolumn{6}{l}{SMC-ECL-2096} \\
\hline
52959.52870   & 204.6    & 0.5               &   78.1   & 0.8               & GIRAFFE                     \\  
52959.70298   &  92.7    & 3.4               &  272.1   & 4.9               & GIRAFFE                     \\  
52960.53008   & 168.2    & 0.8               &  129.9   & 1.7               & GIRAFFE                     \\  
52960.66981   & 252.3    & 1.1               &  -27.3   & 1.1               & GIRAFFE                     \\  
52961.52452   &  70.2    & 1.5               &  294.9   & 1.7               & GIRAFFE                     \\  
52961.70180   & -12.5    & 1.1               &  442.9   & 1.6               & GIRAFFE                     \\  
52962.53471   & 277.5    & 1.0               &  -84.3   & 0.7               & GIRAFFE                     \\  
52962.71513   & 332.3    & 0.6               & -185.2   & 0.9               & GIRAFFE                     \\  
52963.52455   & -13.0    & 0.9               &  452.2   & 1.7               & GIRAFFE                     \\  
52964.52610   & 331.1    & 1.0               & -185.3   & 0.9               & GIRAFFE                     \\  
52964.55559   & 336.3    & 0.6               & -191.4   & 1.1               & GIRAFFE                     \\  
52964.74101   & 311.1    & 0.5               & -137.1   & 0.4               & GIRAFFE                     \\  
52965.51943   & -32.9    & 1.0               &  494.4   & 2.5               & GIRAFFE                     \\
52965.69577   &  19.6    & 0.4               &  397.3   & 0.5               & GIRAFFE                     \\
52966.53717   & 316.1    & 0.7               & -147.6   & 1.4               & GIRAFFE                     \\
52966.69306   & 249.4    & 0.7               &  -18.2   & 1.0               & GIRAFFE                     \\
\hline\noalign{\smallskip}
\multicolumn{6}{l}{SMC-ECL-2073} \\
\hline
52959.52871   & 305.4    & 0.6               & -139.3   & 0.7               & GIRAFFE                     \\
52959.70299   & 232.5    & 1.6               &   -9.8   & 1.7               & GIRAFFE                     \\
52960.53009   &  -3.3    & 0.6               &  416.1   & 1.2               & GIRAFFE                     \\
52960.66982   &  53.3    & 1.2               &  308.2   & 1.5               & GIRAFFE                     \\
52961.52453   & 286.8    & 0.6               & -107.6   & 0.8               & GIRAFFE                     \\
52961.70181   & 202.2    & 1.8               &   48.0   & 1.6               & GIRAFFE                     \\
52962.53472   &  23.8    & 0.8               &  372.3   & 1.3               & GIRAFFE                     \\
52962.71514   & 114.7    & 4.3               &  204.6   & 3.8               & GIRAFFE                     \\
52963.52456   & 261.1    & 0.9               &  -62.7   & 1.1               & GIRAFFE                     \\
52964.52611   &  47.4    & 2.0               &  322.5   & 2.3               & GIRAFFE                     \\
52964.55560   &  66.6    & 1.5               &  295.1   & 2.1               & GIRAFFE                     \\
52964.74102   & 168.3    & 0.3               &  103.1   & 0.6               & GIRAFFE                     \\
52965.51943   & 233.2    & 1.0               &   -9.8   & 1.3               & GIRAFFE                     \\
52965.69578   & 130.7    & 0.7               &  169.1   & 1.3               & GIRAFFE                     \\
52966.53718   &  88.3    & 2.0               &  258.6   & 3.2               & GIRAFFE                     \\
52966.69307   & 176.3    & 3.2               &   94.4   & 4.3               & GIRAFFE                     \\
\hline\vspace{-0.1cm} \\
        \end{tabular}

\begin{tabular}[t]{cccccccc}
\hline
HJD           & RV$_{1}$ & $\sigma_{RV_{1}}$ & RV$_{2}$ & $\sigma_{RV_{2}}$ & \multirow{2}{*}{Instrument} \\
(24\,00000+)  & (\kms)   & (\kms)            & (\kms)   & (\kms)            &                             \\
\hline\noalign{\smallskip}
\multicolumn{6}{l}{SMC-ECL-2213} \\
\hline
52959.52870   & 229.6    & 0.2               &   66.6   & 0.2               & GIRAFFE                     \\ 
52959.70298   & 268.6    & 0.2               &  -65.3   & 0.2               & GIRAFFE                     \\ 
52960.53008   & 182.6    & 0.2               &  129.6   & 0.2               & GIRAFFE                     \\ 
52960.66981   & 138.4    & 0.2               &  273.4   & 0.2               & GIRAFFE                     \\ 
52961.52452   & 102.6    & 0.2               &  348.6   & 0.2               & GIRAFFE                     \\ 
52961.70181   & 162.6    & 0.2               &  196.6   & 0.2               & GIRAFFE                     \\ 
52962.53471   & 289.8    & 0.2               &  -92.3   & 0.2               & GIRAFFE                     \\ 
52962.71513   & 244.6    & 0.2               &   21.6   & 0.2               & GIRAFFE                     \\ 
52963.52455   &  36.6    & 0.2               &  498.6   & 0.2               & GIRAFFE                     \\ 
52964.52610   & 312.6    & 0.2               & -118.3   & 0.2               & GIRAFFE                     \\ 
52964.55559   & 318.6    & 0.2               & -140.3   & 0.2               & GIRAFFE                     \\ 
52964.74101   & 301.6    & 0.2               & -128.3   & 0.2               & GIRAFFE                     \\ 
52965.51943   &  70.1    & 0.2               &  426.6   & 0.2               & GIRAFFE                     \\ 
52965.69577   &  27.7    & 0.2               &  474.6   & 0.2               & GIRAFFE                     \\ 
52966.53717   & 250.6    & 0.2               &   36.6   & 0.2               & GIRAFFE                     \\ 
52966.69306   & 275.6    & 0.2               &  -83.3   & 0.2               & GIRAFFE                     \\ 
\hline\noalign{\smallskip}
\multicolumn{6}{l}{SMC-ECL-1947} \\
\hline
52959.52870   &  34.7    & 4.1               & 295.7    & 5.1               & GIRAFFE                     \\
52959.70297   &  62.7    & 4.1               & 244.7    & 5.1               & GIRAFFE                     \\
52960.53008   & 275.4    & 4.1               & -20.5    & 5.1               & GIRAFFE                     \\
52960.66981   & 268.5    & 4.1               & -21.2    & 5.1               & GIRAFFE                     \\
52961.52452   &  30.6    & 4.1               & 290.6    & 5.1               & GIRAFFE                     \\
52961.70180   &  16.7    & 4.1               & 310.7    & 5.1               & GIRAFFE                     \\
52962.53471   & 210.4    & 4.1               &  70.7    & 5.1               & GIRAFFE                     \\
52962.71513   & 258.0    & 4.1               &   3.0    & 5.1               & GIRAFFE                     \\
52963.52455   & 149.8    & 4.1               & 135.8    & 5.1               & GIRAFFE                     \\
52964.52610   &  74.8    & 4.1               & 232.3    & 5.1               & GIRAFFE                     \\
52964.55558   &  84.6    & 4.1               & 226.6    & 5.1               & GIRAFFE                     \\
52964.74101   & 140.6    & 4.1               & 149.6    & 5.1               & GIRAFFE                     \\
52965.51942   & 267.6    & 4.1               & -17.3    & 5.1               & GIRAFFE                     \\
52965.69577   & 226.5    & 4.1               &  48.6    & 5.1               & GIRAFFE                     \\
52966.53717   &   9.0    & 4.1               & 315.0    & 5.1               & GIRAFFE                     \\
52966.69306   &  25.0    & 4.1               & 307.0    & 5.1               & GIRAFFE                     \\
56567.63244   & 253.2    & 4.1               &   8.0    & 5.1               & GIRAFFE                     \\
56569.59335   & 253.0    & 4.1               &  -2.9    & 5.1               & GIRAFFE                     \\
56587.51531   &  39.1    & 4.1               & 272.4    & 5.1               & GIRAFFE                     \\
56599.71409   &  16.4    & 4.1               & 307.4    & 5.1               & GIRAFFE                     \\
56621.66566   &  49.4    & 4.1               & 276.4    & 5.1               & GIRAFFE                     \\
\hline\noalign{\smallskip}
\multicolumn{6}{l}{SMC-ECL-1903} \\
\hline
52959.52871   & 225.9    & 14.4              &  69.5    & 14.4              & GIRAFFE                     \\
52959.70298   & 211.2    & 10.5              & 103.4    & 10.5              & GIRAFFE                     \\
52960.53009   &  46.6    &  7.8              & 279.1    &  7.8              & GIRAFFE                     \\
52960.66981   &  22.6    &  8.7              & 291.3    &  8.7              & GIRAFFE                     \\
52961.52452   &  55.9    &  9.7              & 271.2    &  9.7              & GIRAFFE                     \\
52961.70181   &  93.2    &  9.8              &   ---    &  ---              & GIRAFFE                     \\
52962.53471   & 294.3    &  9.0              &  -6.3    &  9.0              & GIRAFFE                     \\
52962.71513   & 292.1    &  8.6              & -12.2    &  8.6              & GIRAFFE                     \\
52963.52456   & 230.9    & 14.3              &  60.2    & 14.3              & GIRAFFE                     \\
52964.52610   &  43.5    & 11.8              &   ---    &  ---              & GIRAFFE                     \\
52964.55559   &  38.2    &  8.1              & 291.5    &  8.1              & GIRAFFE                     \\
52964.74101   &   7.1    &  8.6              & 311.1    &  8.6              & GIRAFFE                     \\
52965.51943   &  59.3    &  8.9              & 252.2    &  8.9              & GIRAFFE                     \\ 
52965.69578   & 102.9    &  7.4              &   ---    &  ---              & GIRAFFE                     \\
52966.53718   & 288.5    &  8.7              & -13.5    &  8.7              & GIRAFFE                     \\
52966.69306   & 287.2    &  7.4              &   ---    &  ---              & GIRAFFE                     \\
56599.71409   &  45.3    &  8.5              & 301.5    &  8.5              & GIRAFFE                     \\
56621.66566   & 288.7    & 10.6              &   ---    &  ---              & GIRAFFE                     \\
\hline\noalign{\smallskip}
\multicolumn{6}{l}{SMC-ECL-1105} \\
\hline
56567.63245   &  46.3    & 8.9               & 336.7    &  3.1              & GIRAFFE                     \\
56569.59336   &  82.8    & 8.9               & 237.7    &  6.3              & GIRAFFE                     \\
56587.51531   &  74.7    & 8.9               & 260.7    &  4.2              & GIRAFFE                     \\
56599.71408   & 224.2    & 8.9               & -56.2    &  9.2              & GIRAFFE                     \\
56621.66564   &  49.6    & 8.9               & 316.7    & 11.1              & GIRAFFE                     \\
\hline\vspace{-0.1cm} \\
\end{tabular}
}
\end{center}

\label{lastpage}
\end{document}